\documentclass[english]{IEEEtran}

\usepackage[utf8]{inputenc}
\usepackage{graphics,graphicx,color,epsfig}
\usepackage{amsmath,amssymb,amsthm,amsfonts,bm,bbm}
\usepackage{algorithmicx,algorithm,algpseudocode}
\usepackage{multirow}
\usepackage{cite}
\usepackage[normalem]{ulem}

\usepackage{tikz}
\usetikzlibrary{automata, positioning, arrows,shapes.multipart,fit}
\usepackage{pgfplots}
\usepackage{pgfplotstable}

\definecolor{color00}{HTML}{7CB9E8}
\definecolor{color01}{HTML}{19FF8C}
\definecolor{color02}{HTML}{CC4400}
\definecolor{color03}{HTML}{665500}
\definecolor{color04}{HTML}{0080FF}

\usepackage{subcaption}
\usepackage{todonotes}
\presetkeys{todonotes}{color=blue!20}{}

\pgfplotsset{ignore legend/.style={every axis legend/.code={\renewcommand\addlegendentry[2][]{}}}}

\title{Deep Reinforcement Learning based Blind mmWave MIMO Beam Alignment}
\author{Vishnu Raj \hspace{8pt} Nancy Nayak \hspace{8pt} Sheetal Kalyani\\
   \hspace{-0 cm}Department of Electrical Engineering, \\ Indian Institute of Technology Madras, \\
   Chennai, India, 600 036. \\
   \texttt{\{ee14d213@ee,ee17d408@smail,skalyani@ee\}.iitm.ac.in}
}

\begin{document}
	\maketitle

	\begin{abstract}
		Directional beamforming is a crucial component for realizing robust wireless communication systems using millimeter wave (mmWave) technology. Beam alignment using brute-force search of the space introduces time overhead while location aided blind beam alignment adds additional hardware requirements to the system. In this paper, we introduce a method for blind beam alignment based on the RF fingerprints of user equipment obtained by the base stations. The proposed system performs blind beam alignment on a multiple base station cellular environment with multiple mobile users using deep reinforcement learning. We present a novel neural network architecture that can handle a mix of both continuous and discrete actions and use policy gradient methods to train the model. Our results show that the proposed method can achieve a data rate of up to four times the traditional method without any overheads.
	\end{abstract}
	
	\section{Introduction}
Millimeter Wave (mmWave) systems are considered as one of the key technologies in next-generation wireless communication systems. Massive MIMO systems with mmWave technologies combine the advantages of leveraging spatial resources of MIMO with the high data rates offered by the large bandwidth available at millimeter wave frequency bands. However, excessive pathloss and penetration loss incurred during wave propagation severely affect the range of mmWave MIMO systems. Directional transmission and antenna beamforming have been proposed as a solution for compensating for the losses incurred during wave propagation \cite{kutty2015beamforming}.

Directional beamforming has been traditionally achieved through beam sweeping methods \cite{nitsche2014ieee}, which involves a brute-force search through the possible steering directions. But brute-force search method is both time-consuming and requires energy expenditure. To reduce the complexity of the full-scale search, hierarchical search strategies have been proposed (See \cite{zhang2017codebook} and references therein). Blind beam steering relying on accurate location information has been proposed as a low complexity solution for beamforming \cite{nitsche2015steering}, but comes with the overhead of additional location information. In recent work, \cite{liu2019ekf} develop a method based on Extended Kalman Filter to track the mmWave beam in a mobile scenario with moving UEs. In \cite{zhang2019exploring}, a temporal correlation aided beam alignment scheme is presented. By describing the transition probabilities of Angle-of-Arrival (AoA) and Angle-of-Departure (AoD), the authors developed a semi-exhaustive search algorithm for beam alignment. \cite{yan2020fast} presents a solution to improve the Initial Access (IA) process of beam alignment in High-Speed-Railways (HSR). By exploiting the periodicity and regularity in the train trajectory and using historical beam training results, the authors proposed a method to reduce IA. A Genetic algorithm based search method for Initial Access is proposed in \cite{souto2019novel}. In the context of 5G millimeter wave communication, the authors showed that the proposed method can achieve the same capacity as exhaustive search algorithms under small parameter settings. However, even with better IA schemes, there are chances of beam misalignment and precise tracking may be a costly alternative. The work in \cite{pradhan2019beam} developed a beam misalignment aware hybrid precoding scheme to deal with the mismatch in estimation. By using GPS location data from UE and querying a multipath fingerprint database of historical data, \cite{va2017inverse} introduced a beam alignment method for vehicular UEs.

The progress in deep learning in the areas of computer vision and speech signal processing\cite{lecun2015deep} has also triggered an interest in applying those techniques to complex wireless communication problems \cite{zhang2019deep}. Deep learning has been successfully used in channel estimation \cite{ye2018power,athreya2020beyond}, end to end communication systems \cite{o2017introduction,raj2018backpropagating,raj2020design}, OFDM systems \cite{gao2018comnet} etc. The application of developments in deep learning research has also improved the solutions for mmWave communication systems. Some of the fruitful applications include beamforming design for weighted sum-rate maximization \cite{huang2018unsupervised}, using an autoencoder deep learning model to improve hybrid precoding \cite{huang2019deep}, replacing hybrid precoding with a deep learning model to predict the best pair transmit/receive beam pairs from the observed channel\cite{li2019deep} and leveraging deep reinforcement learning for beamforming \cite{wang2020precodernet}. In the scenario of hybrid beam forming, \cite{elbir2019joint} proposed a Convolutional Neural Network based method for joint antenna selection and beamforming using twin networks. It can be observed from these works that mmWave MIMO systems can greatly benefit from the application of learning techniques into their core components.

\begin{figure}[h]
    \centering
    \input{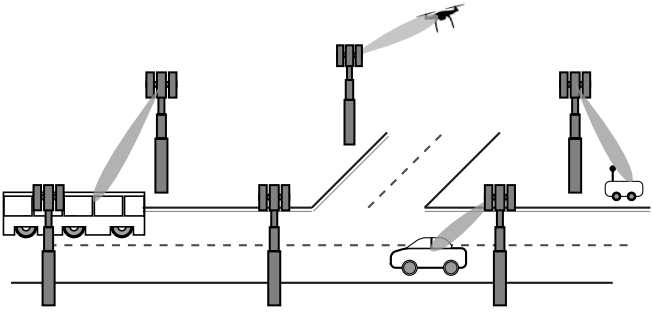}
    \caption{A depiction of problem scenario}
    \label{fig:city_block}
\end{figure}

In this work, we consider the problem of blind mmWave beam alignment in a downlink channel using only the RF signature about the presence of UE in the cellular system. Specifically, we consider a multi-base station (BS) scenario with multiple mobile users (UE). A simple depiction of the scenario is given in Fig. \ref{fig:city_block}.

We consider a scenario where multiple mmWave micro base stations ($\mu$BS) exist and there is a central base station (not depicted in Fig. \ref{fig:city_block}) which co-ordinates all the transmissions through $\mu$BS\footnote{ The learning algorithm presented in this work runs at the central base station which in turn decides the UE-$\mu$BS assignment and also the beam alignment directions based on the assignment. Extending to distributed learning through Federated Learning can be interesting future work.}. Due to the path loss and penetration loss properties of mmWave, each $\mu$BS has only a limited coverage area, usually within hundreds of meters. As the mobile users move in the environment, we need to select the best $\mu$BS to serve each user based on the channel characteristics between $\mu$BS and users. These user equipments can be any mobile transceiver that employs mmWave spectrum for communication. Examples of such user terminals include mobile phones, connected vehicles, autonomous cars, delivery robots, unmanned aerial vehicles (UAVs), etc. Specifically, our aim is to select one $\mu$BS out of the available base stations to serve each user and also to find the best beam alignment angles for transmission without any brute force beam sweep methods. Rather than relying on the location information for beam alignment as considered in previous works, the proposed method uses already available radio frequency (RF) signature available in the system about the presence of the UEs and leverage the advancements in deep reinforcement learning to achieve blind beam alignment from BS to UE for downlink communication.

\textbf{Related Works.} Deep learning-based solutions for beam alignment with no location information have been proposed in multiple scenarios. In \cite{alkhateeb2018deep}, a coordinated beamforming solution using deep learning to enable high mobility and high data rate is proposed. Based on the uplink pilot signal received at the terminal BSs, a deep neural network is trained to predict the best beamforming vectors. The method is applied to the scenario where multiple BSs serve a single UE. Extending this method to support multiple UEs may not control the interference between the UEs as the existence of multiple UEs is not known to the trained deep learning model. Another approach that takes into account the varying traffic patterns and its impact on the mmWave channel is proposed in \cite{satyanarayana2019deep}. By taking the location, traffic parameters and RSSI thresholds of each UE, the proposed method suggests a set of beamforming vectors based on an estimated RF fingerprint which can then be used to conduct beam training. By deploying a deep learning based method to shortlist possible best beamforming vectors, this method can reduce the beam training time for initiating a mmWave communication. However, to train this model, an exhaustive dataset of possible RF fingerprints across multiple traffic patterns is required. On the other side, once trained, the model can continue to operate as long as there is no major change in the environment. In an alternative take on the problem, \cite{klautau2019lidar} proposed a solution that includes BSs broadcasting its location. Here, UEs (connected vehicles with LIDARs) fuses the information from its LIDAR and the received BS location information to shortlist a possible set of beamforming vectors. This also helps in reducing the delay due to brute force beam search. The work in \cite{dias2019position} proposed to use the LIDAR in connected smart vehicles such as autonomous driving vehicles to estimate the location of BS and then aiding beamforming. Instead of BS broadcasting its location, this method allows smart vehicles with LIDARto detect the location of BS through its LIDAR scans and complement this data with deep learning to reduce the search space for beamforming vectors. 

All these works mentioned above relies on additional resources such as GPS hardware to acquire location information\cite{nitsche2015steering,va2017inverse}, LIDAR hardware to acquire contextual information, etc \cite{klautau2019lidar,dias2019position}, for beam alignment. Even though all these solutions are able to present improved performance, additional hardware/resource requirements are a hindrance to the widespread adoption of these solutions. Further, human aided acquisition of labeled dataset for training deep learning models severely limit the scalability of the solutions.

\textbf{Contributions.} In this work, we propose a deep reinforcement learning \cite{sutton2018reinforcement, mnih2015human} based technique for blind beam alignment that does not need any additional hardware/resources and does not require any labeled dataset for training. The proposed method is designed to work in a scenario with multiple base stations as well as multiple UEs. By using the RF fingerprint of each UE produced by its omnidirectional beacon transmission\footnote{Even though an explicit mention of beacon signaling is assumed in this work, this beacon signal from UE can be any of the existing control signals which are required to maintain connectivity between UE and BS as the work only requires the received signal strength of the beacon signal and not the specific signals themselves.}, the system learns to predict the best BS to serve each UE as well as the beam alignment directions for each transmission. This makes the proposed solution also ideal for situations where UEs and environments are non-stationary since in such a case any contextual information obtained about the location alone may not be useful for beamforming. 
The numerical experiments performed in the paper shows that the proposed method not only increases the average sum rate of UE four times when compared to the traditional beam-sweeping method in the case of small antenna arrays but also reaches the same rate as an Oracle that has full information about the environment. Also, the average sum rate of the proposed method is remarkably better than the corresponding rate values of traditional methods like BS-sweep or the DRL based methods like Vanilla DDPG even in the case of larger antenna arrays.
As the proposed method is not dependent on the explicit channel modeling, it can be used for both ground-based as well as aerial communication networks.

\textbf{Notations.} Bold face lower-case letters (eg. $\bm{x}$) denote column vector and upper-case (eg. $\bm{M}$) denote matrices. Script face letters (eg. $\mathcal{S}$) denotes a set, $|\mathcal{S}|$ denotes the cardinality of the set $\mathcal{S}$. $f(\textbf{x};\bm{\theta})$ represents a function which takes in a vector $\textbf{x}$ and has parameters $\bm{\theta}$. A distribution with parameters $\theta$ is represented as $p_\theta(\cdot)$. $\mathbb{E}_p$ is the expectation operator with respect to distribution $p$.

	\section{Blind mmWave wave Beam Alignment}
We consider a mmWave downlink scenario with multiple BSs trying to serve multiple 
UEs. The aim is to select the best BS to serve each UE as well as the best set of 
beam alignment parameters for efficient beamforming. This is achieved by the central
base station selecting one $\mu$BS which can best serve the UE based on the current
information about UE (the received signal strength of beacon signal from UEs) and 
then selecting the right beam alignment directions. Each base station has a small 
cell radius (hundreds of meters) within which it can serve and there exists a central 
base station (BS) that coordinates the cellular system. We assume that a reliable 
link exists between the micro base stations ($\mu$BS) and the central BS which can 
ensure a robust exchange of data. For any additional signaling, we assume the existence 
of a dedicated control channel (CC). Also, we consider that all UEs use the same 
carrier frequency and hence the interference should also be reduced while selecting 
the beamforming parameters. 

\subsection{Channel Model}
Let $N_{BS}$ represents the number of $\mu$BS and $N_{UE}$ represents the number
of UEs. We consider a MISO system with $N_T$ transmit antennas at each $\mu$BS and 
$N_R = 1$ receive antenna at each UE. With single antenna UEs, have omni-directional 
transmission. We consider a Uniform Planer Array (UPA) antenna at $\mu$BS. The channel 
between the $\mu$BS and UE (of dimension $N_R \times N_T$) is modeled based on the 
popular Saleh-Valenzuela channel model \cite{el2014spatially} for mmWave systems. 
The channel between the transmitter and receiver is given by
\begin{align}
    \bm{H} = \sqrt{\frac{N_T N_R}{\kappa}}
                \sum \limits_{l=1}^{\kappa}
                    \alpha_l
                        \bm{a}_r(\phi_l^r,\theta_l^r)
                        \bm{a}^*_t(\phi_l^t,\theta_l^t),
\end{align}
where $\kappa$ is the number of propagation paths, $\alpha_l$ is the complex gain 
associated with $l^{th}$ path. $(\phi_l^t,\theta_l^t)$ are the azimuth and elevation 
angles of departure of $l^{th}$ ray at the $\mu$BS respectively. Similarly, $(\phi_l^r,
\theta_l^r)$ are the angles of arrival of $l^{th}$ ray at UE. We measure $\theta$ 
from \textit{+z-axis} and $\phi$ from \textit{+x-axis}. We assume the $\mu$BS UPA 
antenna is in \textit{yz-plane} with $N_{t,h}$ and $N_{t,v}$ elements in $y$ and 
$z$ axis respectively and $N_t = N_{t,h} \times N_{t,v}$. The array response vector 
of transmitter, $\bm{a}_{t}(\phi,\theta)$ (of dimension $N_T$), is given by
\ifCLASSOPTIONonecolumn
\begin{align}
        \bm{a}_{t}(\phi,\theta) = \frac{1}{\sqrt{N}} \left[ 
            1, 
            \ldots,
            e^{j \frac{2\pi}{\lambda} d \left(m \sin\phi \sin\theta + n \cos \theta \right)},
            \ldots, 
            e^{j \frac{2\pi}{\lambda} d \left((N_{t,h}-1) \sin\phi \sin\theta 
                    + (N_{t,v}-1) \cos \theta \right)} \label{eqn:resp_vec}
        \right]^T
\end{align}
\else
\begin{align}
    \bm{a}_{t}(\phi,\theta) = \frac{1}{\sqrt{N}} \left[ 
            1, 
            \ldots,
            e^{j \frac{2\pi}{\lambda} d \left(m \sin\phi \sin\theta + n \cos \theta \right)},
            \ldots, 
            \right. \nonumber \\ \left.
            e^{j \frac{2\pi}{\lambda} d \left((N_{t,h}-1) \sin\phi \sin\theta 
                    + (N_{t,v}-1) \cos \theta \right)} \label{eqn:resp_vec}
        \right]^T
\end{align}
\fi
where $0 < m < N_{t,h} - 1$ and $0 < n < N_{t,v} - 1$ and $d$ is the inter-element spacing.
Since we assume omni-directional reception with single antenna at the receiver, 
$\bm{a}^{r}(\phi,\theta) = 1$. The path loss (in $dB$) of mmWave propagation is 
modeled as \cite{5gmodel2016},
\ifCLASSOPTIONonecolumn
\begin{align}
        PL(f,d)_{dB} &= 20 \log_{10} \left( \frac{4\pi f}{c} \right) 
        + 10 n \left(1 + b \left(\frac{f-f_0}{f_0}\right) \right)
        \log_{10} \left(d\right)
        + X_{\sigma dB}, \label{eqn:pathloss}
\end{align}
\else
\begin{align}
    PL(f,d)_{dB} &= 20 \log_{10} \left( \frac{4\pi f}{c} \right) 
                \nonumber \\
                & \qquad + 10 n \left(1 + b \left(\frac{f-f_0}{f_0}\right) \right)
                    \log_{10} \left(d\right) \nonumber \\
                & \qquad + X_{\sigma dB}, \label{eqn:pathloss}
\end{align}
\fi
where $n$ is the path loss exponent, $f_0$ is the fixed reference frequency, $b$ 
captures the frequency dependency of path loss exponent and $X_{\sigma dB}$ is the 
shadow fading term in $dB$.

The Signal to Interference Ratio (SINR) at $i^{th}$ UE which is served by $j^{th}$ 
$\mu$BS is given by
\begin{align}
    \zeta^{i} = \frac{P_{TX}|\bm{H}_{i,j} \bm{f}_j|^2}
                {\sum \limits_{k=1, k \neq j}^{N_{BS}} P_{TX,i}|\bm{H}_{i,k} \bm{f}_k|^2
                    + \sigma^2}, \label{eqn:ue_sinr}
\end{align}
where $\bm{H}_{i,j}$ is the channel between $i^{th}$ UE and $j^{th}$ BS, $\bm{f}_j$
is the transmit codeword used by $\mu$BS, $\sigma^2$ is the noise power and $P_{TX,k}$
is the transmit power of $k^{th}$ $\mu$BS. The transmit codeword $\bm{f}_j$ is computed 
from beamforming angles $(\phi_{ij}, \theta_{ij})$ as $\bm{f}_j = \bm{a}(\phi_{ij}, 
\theta_{ij})$ as given in (\ref{eqn:resp_vec}).

\textbf{RF signature of UE.} Each $UE$ transmits a uniquely identifiable beacon
signal using omnidirectional transmission during the period of downlink. All $\mu$BS 
in the system receives a faded/corrupted copy of this uniquely identifiable signal. 
Since there may not exist direct pathways from the UE to BS for these beacon signals, 
the received power at each BS can have a complex relationship that is dictated by 
the scenario. The final RF signature used by the learning system is a vector of $N_{BS}$ 
dimensions, which is the received signal strength of the UE beacon signal at each 
of the $N_{BS}$ micro base stations.

The learning problem is now to use the RF signatures along with the reported SINR
values of each UE to predict which $\mu$BS should serve which UE and with what
values of $\phi$ and $\theta$. 
An outline of the entire procedure is given below.
\begin{enumerate}
    \item Each UE send out a uniquely identifiable beacon signal using omnidirectional
            transmission.
    \item Each $\mu$BS receives the beacon signals transmitted by all the UEs. These
            received power of the beacon signals at each $\mu$BS constitutes the 
            RF signature for UEs.
    \item Central base station receives the RF signature collected by $\mu$BS about
            each UE. Central base station also obtains the SINR from each UE about
            the ongoing downlink transmission.
    \item Based on the RF signature and the SINR of each UE, the central base station
            runs the proposed algorithm and selects the best $\mu$BS for each UE and
            the beam alignment angles.
    \item Central base station commands the selected $\mu$BS to use the predicted
            angle parameters and serve particular UE.
    \item Corresponding $\mu$BS performs beam alignment based on the received command
            from the central base station for transmission until an update.
\end{enumerate}

This process can be repeated periodically or in an event-triggered fashion to update
the beam alignment process for each UE. Compared to the standard procedure of beam 
sweeping which does brute force search in all available directions for the best channel
response, the proposed approach can get the beam alignment parameters in a single shot 
rather than searching for multiple combinations. Further, this method also enables 
the central base station to choose the best $\mu$BS for each UE based on the instantaneous
SINR feedback and hence improves the initial access problem as well.

In the following section, we model the problem of selecting $\mu$BS and the beam 
alignment direction as a single Markov Decision Process (MDP) with the objective 
of improving the effective SINR experienced by each of the UEs. We then introduce 
Reinforcement Learning as a viable solution to solve the MDP problem and then proceed 
with the specific neural architecture as well as the learning algorithm for the problem.

	\section{Learning based beam alignment}
\begin{figure}[h]
    \centering
    \tikzstyle{block} = [rectangle, draw, 
    text width=8em, text centered, rounded corners, minimum height=4em]
    
\tikzstyle{line} = [draw, -latex]

\begin{tikzpicture}[node distance = 6em, auto, thick, scale=0.6]
    \node [block] (BS) {Central Basestation};
    \node [block, below of=BS] (muBS) {$\mu$BS};
    \node [block, below of=muBS] (UE) {User Equipments};
    
     \path [line] (BS.0) --++ (4em,0em) |- node [near start]{$a_t$} (muBS.0);
     \path [line] (UE.180) --++ (-6em,0em) |- node [near start] {$r_{t}$} (BS.170);
     \path [line] (muBS.180) --++ (-4.25em,0em) |- node [near start, right] {$s_{t+1}$} (BS.190);
     \path [line] (muBS.280) -- node [right] {beam} (UE.80);
     \path [line] (UE.100) --node [left] {beacon} (muBS.260);
\end{tikzpicture}
    \caption{Blind beam alignment using DRL.}
    \label{fig:mmbeamer}
\end{figure}
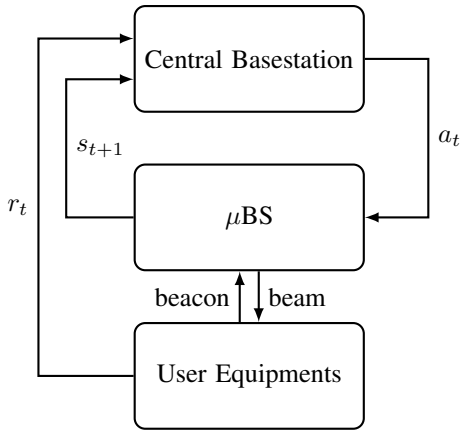

We model the problem of $\mu$BS selection and beam alignment direction prediction
as a Markov Decision Process (MDP). An MDP comprises of a state space $\mathcal{S}$, 
an action space $\mathcal{A}$, an initial state distribution $p(s_1)$, a stationary 
distribution for state transition
which obeys Markov property $p(s_{t+1}|s_{t},a_{t}) = p(s_{t+1}|s_{t},a_{t}, \ldots,
s_{1},a_{1})$ and a reward function $r: \mathcal{S} \times \mathcal{A} \rightarrow
\mathbb{R}$. 
An overview of the MDP formulation for the problem of blind beam alignment is given 
in Fig. \ref{fig:mmbeamer}.

The input to the learning agent is a $N_{UE} \cdot (N_{BS} + 1)$ dimensional vector
with real number entries
which is a concatenation of $N_{BS} + 1$ feature vector of $N_{UE}$ UEs in the network.
The feature vector of dimension $N_{BS} + 1$ of each UE consists of a) the RF fingerprints 
($N_{BS}$ elements) from seen at $\mu$BS about the $UE$ from the received beacon 
signal and b) the SINR experienced by that UE with current beam alignment configuration 
and from the $\mu$BS assigned to the UE. At each time step $t$, this 
constitutes the state
described in the MDP ie., $s_t \in \mathbb{R}^{N_{UE} \cdot (N_{BS} + 1)}$. Action
to be taken by the agent has to include a discrete value referring to the index of
the $\mu BS$ to serve each UE and also the $(\phi,\theta)$ pair for the transmission.
The reward at each time instant $t$, $r_t$ is taken as the mean rate achieved by 
the UEs at that instant and is defined as,
\begin{align}
    r_t &= \frac{1}{N_{UE}} \sum \limits_{i=1}^{N_{UE}} \log_2 
                \left( 1 + \zeta_t^i \right),
                \label{eqn:reward}
\end{align}
where $\zeta_t^i$ is the instantaneous SINR at $i^{th}$ $UE$ as defined in 
(\ref{eqn:ue_sinr}).

\subsection{Reinforcement Learning}
Reinforcement Learning (RL) \cite{sutton2018reinforcement} is a sub-class of artificial
intelligence (AI) where the learning algorithm \textit{interacts} with the \textit{environment}
to learn \textit{optimal actions} which maximizes a formulated \textit{reward}.
It differs from the Machine Learning (ML) paradigm in the way how samples are
obtained for the learning process. Machine learning relies on labeled data to
be fed into the algorithm for the learning process, reinforcement learning works by
the learning agent itself acquiring the samples to improve its knowledge. Machine 
learning is ideal for scenarios where labeled data is available such as in classification 
and regression, and where the prediction itself is not going to affect the 
future observations. Reinforcement learning is used in scenarios where the learning
agent is in control of the system whose output can change based on the agent's predictions.
This difference also creates an \textit{explore-exploit} behavior in RL algorithms
where it is also required to explore unknown/less-known actions to acquire new 
samples.

In reinforcement learning, an agent is trained to optimize a policy $\bm{\pi}$ to increase
the return $\mathbbm{r}_t(\gamma)$. 
A policy $\bm{\pi}$ maps observed states to actions $\pi : \mathcal{S} \rightarrow
\mathcal{A}$ and inturn obtains reward $r_t(s_t, a_t)$.
The return $\mathbbm{r}_t(\gamma)$ is defined
as the total discounted reward from the timestep $t$ and can be expressed as
$
    \mathbbm{r}_t(\gamma) = \sum \limits_{t'=t}^{\infty} 
                                        \gamma^{t'-t} r(s_{t'},a_{t'}),
                                        \label{eqn:return}
$
where $\gamma \in [0,1]$. The discounting factor $\gamma$ is used to capture the
importance of future rewards in the current value estimate. With $\gamma \rightarrow 
0$, the policy will become myopic and only considers current reward. With $\gamma
\rightarrow 1$, the policy learns for long-term high reward.
The objective of agent is to find a policy $\bm{\pi}$ which 
maximizes the expected cumulative discounted return $J(\bm{\pi}) = \mathbb{E}[\mathbbm{r}_1(\gamma)
|\bm{\pi}]$. An agent computes the value function by a policy $\bm{\pi}$ for 
each state as the expected return from that state by following $\bm{\pi}$, ie., $V^\pi(s)
= \mathbb{E}\left[\mathbbm{r}_1(\gamma)| S_1 = s; \bm{\pi} \right]$. The value of a state indicates
how favorable each state is for the agent to be in. The quality of an action at each
state is computed using Q-function as $Q^\pi(s,a) = \mathbb{E}\left[\mathbbm{r}_1(\gamma)|
S_1 = s, A_1 = a; \bm{\pi} \right]$ and indicates how rewarding each action is when 
taken from that state $s$. At each timestep, agent takes action which maximizes 
the Q-value. 

The policy $\bm{\pi}$ can either be created
by a tabular approach or by function approximation approach\cite{sutton2018reinforcement}.
In the case where the states and actions are discrete, tabular methods can be used
to capture the value for $Q^\pi(s,a)$, which subsequently decides the action taken
by policy $\bm{\pi}$. On the other hand, when states/actions can take continuous values
or there are very high number of states/actions which can make maintaining a table for
Q-values infeasible, function approximation methods are used. By appropriately defining
a family of functions that operate on $(s,a)$ and learning the parameters of the function
which maximizes the return from the policy $\bm{\pi}$, these methods depend on the
expressive power of candidate functions for the success of RL policies. Multiple
function approximation techniques including linear functions, radial basis functions,
Fourier basis, neural networks, etc has been proposed \cite{konidaris2011value}.


\subsection{Deep Reinforcement Learning based Beam Alignment}
Deep Reinforcement Learning (DRL) \cite{mnih2015human} is the technique where  
an RL learning agent employs deep neural networks for facilitating the learning
procedure.
The function approximation capabilities of the neural networks are
leveraged in DRL to map the observations to optimal actions. A neural network
can be seen as a chain of functions that transforms its input to a set of outputs through a non-linear transform. In DRL, at each time step, after observing the state $\bm{s}$, agent uses a neural network policy $\bm{\pi}$ to take an action $\bm{a}$. 

The problem of beam alignment as formulated above is a challenging task for reinforcement learning as the action space $\mathcal{A}$ is a mixture of continuous (the value of angles) as well as discrete (the index to $\mu BS$) dimensions. Deep Deterministic Policy Gradient (DDPG) algorithm \cite{lillicrap2015continuous} is a recent actor-critic method for training deep reinforcement learning agents on continuous action domains. 
In this work, we use DDPG as the learning algorithm to train DRL agent\footnote{Even 
though we use DDPG in this work, any DRL method with continuous action space can 
be used with the proposed neural action predictor for the purpose of blind beam alignment}.
As our problem has an action space which is a mix of discrete base station
selection and continuous beam alignment angles selection, we propose a novel neural
network architecture that can handle pseudo-discrete and pure-continuous action
spaces simultaneously for predicting actions.

\subsubsection{\textbf{Training the agent with DDPG}}
DDPG is based on the family of policy gradient
    algorithms in which the parameters of the policy are changed towards the direction
    of improvement of return. It is a two-step iterative process in which the policy is
    evaluated for the quality with current set of parameters and then, a policy improvement
    step updates the parameters in the ascent direction of maximum returns. DDPG has two neural networks: an actor network $\mathbbm{A}$ parameterized by $\bm{\omega}^a$ which predicts the action $a_t$ based on current state $s_t$ and critic network $\mathbbm{C}$ parameterized by $\bm{\omega}^c$ which computes the Q-value for the predicted action $Q(s_t, a_t)$. The critic network $\mathbbm{C}$ predicts the quality of the action taken by actor network $\mathbbm{A}$ and hence encourages the actor network to take \textit{better} actions through its feedback. The critic network $\mathbbm{C}$ meanwhile trains itself for better prediction by observing the rewards after each action and hence computing the Q-value followed by gradient of the error on its prediction. To get stable, uncorrelated gradients for policy improvement, DDPG maintains a replay buffer of finite size $\tau$ and sample the observations from the buffer in minibatches to update the parameters. DDPG also uses target networks with parameters $\bar{\bm{\omega}}^a$ and $\bar{\bm{\omega}}^c$ to avoid divergence in value estimation. This helps the learning agent to update the parameters of the active network based on the values from the target network, hence giving the learning agent a stable error value to learn from. At each timestep, the state $s_i$ and the action takes $a_i$ along with the reward obtained $r_i$ and the next state $s_{i+1}$ is stored as an experience $(s_i, a_i, r_i, s_{i+1})$ to the buffer $\mathcal{B}$. For training the actor and critic networks, $N$ samples are taken from $\mathcal{B}$ and this is used to compute the gradients. For the critic network $\mathbbm{C}(\bm{\omega}^c)$ to compute the Q-value for each state action-pair, an estimate of return for state $s_i$ in each sample is computed as
    \begin{align}
        y_i = r_i + \gamma \mathbbm{C}(s_{i+1}, \mathbbm{A}(s_{i+1}|
                                \bar{\bm{\omega}}^a)|\bar{\bm{\omega}}^c).
                                \label{eqn:critic_return}
    \end{align}
    Based on the estimate for return, the Mean Squared Bellman Error (MSBE) is computed as
    \begin{align}
        \mathcal{L} = \frac{1}{N} \sum \limits_{i} \left( 
                y_i - \mathbbm{C}(s_i, a_i|\bm{\omega}^c)
            \right)^2. \label{eqn:critic_msbe}
    \end{align}
    Then, the critic network parameters are updated as
    \begin{align}
        \bm{\omega}^c \gets \bm{\omega}^c - \eta_c \nabla_{\bm{\omega}^c} \mathcal{L},
            \label{exp:critic_update}
    \end{align}
    where $\eta_c << 1$ is the stepsize for stochastic update. For the actor network, the update depends on both the gradient of action as well as the improvement in Q-value. The final update for updating parameters of critic network $\bm{\omega}^c$ is given by
    \begin{align}
        \bm{\omega}^a \gets \bm{\omega}^a + 
                            \eta_a \frac{1}{N} \sum \limits_{i} 
                            \left( \nabla_{\bm{\omega}^a} \mathbbm{A}(s) 
                                \nabla_a \mathbbm{C}(s,a) | _{a=\mathbbm{A}(s)} \right),
            \label{exp:actor_update}
    \end{align}
    where $\eta_a << 1$ is the update stepsize. Finally, the target network parameters are updated in every timestep to provide stable value estimates using an exponentially weighted update as
    \begin{align}
        \bar{\bm{\omega}}^c &\gets \lambda \bm{\omega}^c
                            + (1 - \lambda) \bar{\bm{\omega}}^c;  \:
                        \bar{\bm{\omega}}^a \gets \lambda \bm{\omega}^a
                            + (1 - \lambda) \bar{\bm{\omega}}^a,
            \label{exp:target_update}
    \end{align}
    with $\lambda << 1$. Interested readers are directed to  \cite{silver2014deterministic,lillicrap2015continuous} for more information.

\subsubsection{\textbf{Architecture of Neural Action predictor of the proposed method}}
DDPG is originally proposed for continuous action spaces. Since the problem of $\mu BS$ selection is discrete and selecting $(\theta,\phi)$ is continuous, a direct application of DDPG for the problem will result in an inefficient learning procedure as no information about the discrete part of the action is taken into account\footnote{We provide the results for using Vanilla DDPG to our problem in Results section.}. Hence, we propose a novel architecture for neural function approximators that can be used for both discrete and continuous action spaces.

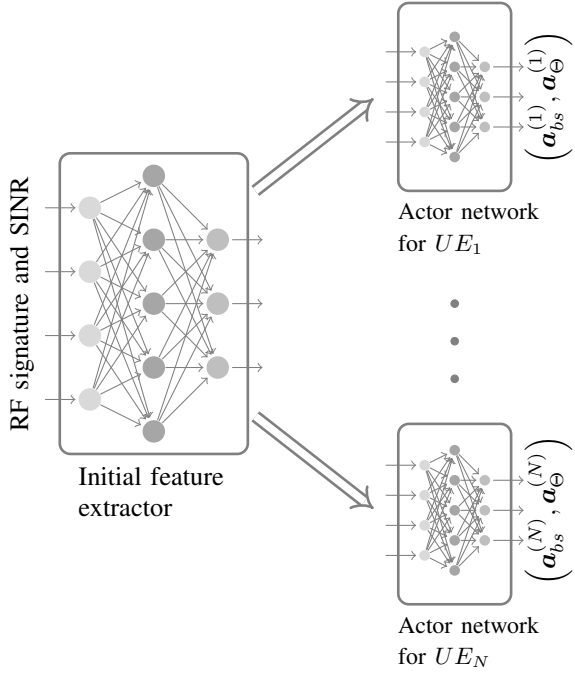
\begin{figure}[h]
        \centering
        \tikzset{
    pics/ann/.style={
        code={
            \tikzstyle{every pin edge}=[<-,shorten <=1pt]
            \tikzstyle{neuron}=[circle,fill=black!25,minimum size=#1*10pt,inner sep=0pt]
            \tikzstyle{input neuron}=[neuron, fill=gray!30, pin=left:];
            \tikzstyle{output neuron}=[neuron, fill=gray!50, pin={[pin edge={->}]right:}];
            \tikzstyle{hidden neuron}=[neuron, fill=gray!70];
            
            \node[input neuron] (I1) at (-1.0*#1, -1.5*#1) {};
            \node[input neuron] (I2) at (-1.0*#1, -0.5*#1) {};
            \node[input neuron] (I3) at (-1.0*#1, +0.5*#1) {};
            \node[input neuron] (I4) at (-1.0*#1, +1.5*#1) {};

            \node[hidden neuron] (H1) at (0.0*#1, -2.0*#1) {};
            \node[hidden neuron] (H2) at (0.0*#1, -1.0*#1) {};
            \node[hidden neuron] (H3) at (0.0*#1, +0.0*#1) {};
            \node[hidden neuron] (H4) at (0.0*#1, +1.0*#1) {};
            \node[hidden neuron] (H5) at (0.0*#1, +2.0*#1) {};

            \node[output neuron] (O1) at (1.0*#1, -1.0*#1) {};
            \node[output neuron] (O2) at (1.0*#1, +0.0*#1) {};
            \node[output neuron] (O3) at (1.0*#1, +1.0*#1) {};

            \foreach \source in {1,...,4}
                \foreach \dest in {1,...,5}
                    \path (I\source) edge (H\dest);
            
            \foreach \source in {1,...,5}
                \foreach \dest in {1,...,3}
                    \path (H\source) edge (O\dest);
        }
    }
}

\tikzstyle{block} = [rectangle, draw, 
    text width=8em, text centered, rounded corners, minimum height=4em]

\begin{tikzpicture}[shorten >=1pt,->,draw=black!50]

    
    \draw[rounded corners,line width=1pt] (1.0,2) rectangle (3.5,6);
    \pic at (2.25,4.0) {ann=0.85};
    \node[rotate=90] (input) at (0.5,4) {RF signature and SINR};
    \node[text width=2cm] (initial) at (2.25,1.50) {Initial feature extractor};
    
    \draw[rounded corners,line width=1pt] (5.5,0) rectangle (7.0,2.4);
    \pic at (6.25,1.25) {ann=0.40};
    \node[rotate=90] (ue1_out) at (7.5,1.25) {$\left(\bm{a}_{bs}^{(N)}, \bm{a}_{\Theta}^{(N)}\right)$};
    \node[text width=2cm] (initial) at (6.50,-0.50) {\small Actor network for $UE_N$};
    
    \draw[rounded corners,line width=1pt] (5.5,5.5) rectangle (7.0,8.0);
    \pic at (6.25,6.75) {ann=0.40};   
    \node[rotate=90] (ueN_out) at (7.5,6.75) {$\left(\bm{a}_{bs}^{(1)}, \bm{a}_{\Theta}^{(1)}\right)$};
    \node[text width=2cm] (initial) at (6.50,5.00) {\small Actor network for $UE_1$};

    \draw[line width=1pt,-implies,double, double distance=1mm] (3.6,2.5) -- (5.2,1.25);
    \draw[line width=1pt,-implies,double, double distance=1mm] (3.6,5.5) -- (5.2,6.75);

    \node[draw,shape=circle,fill=gray,minimum size=1mm,inner sep=0pt] (d1) at (6.25,4.00) {};
    \node[draw,shape=circle,fill=gray,minimum size=1mm,inner sep=0pt] (d2) at (6.25,3.50) {};
    \node[draw,shape=circle,fill=gray,minimum size=1mm,inner sep=0pt] (d3) at (6.25,3.00) {};
\end{tikzpicture}
        \caption{Architecture of Proposed Neural Action Predictor.}
        \label{fig:nn_arch}
    \end{figure}

The purpose of critic network $\mathbbm{C}$ is to predict the estimate the Q-value for each state-action pair. As Q-value is continuous, a traditional feedforward neural network with a scalar output can be used as $\mathbbm{C}$, as used in DDPG. However, the actor network is responsible for predicting the best action $a_t$ given a state $s_t$ and the neural function approximator for $\mathbb{A}$ needs to handle both discrete and continuous spaces. In the proposed architecture for actor network, we split the predictions for each UE through a sub-network at the output. All sub-networks share a common feature extractor which operates on the input to provide each UE sub-networks with a set of features that can be used to select the action corresponding to that UE. The architecture of the proposed Neural Action Predictor is given in Fig. \ref{fig:nn_arch}.

At timestep $t$, let $\bm{x}_0 = s_t$ be the input to the common feature extractor network. We have $s_t \in \mathbb{R}^{N_{UE}\cdot(N_{BS}+1)}$. The first $L$layers of the actor network constitute the common feature extractor network. At each layer, a linear combination of features from previous layer is created and is then passed through a non-linear activation function. Let $\bm{W}_l \in \mathbb{R}^{d_{l,o} \times d_{l,i}}$ and $\bm{b}_l \in \mathbb{R}^{d_{l,o}}$ be the weight and bias associated with layer $l$ and $d_{l,i}$ and $d_{l,o}$ be the input and output dimensions of $l^{th}$ layer. The output $\bm{x}_l$ from $l^{th}$ feature extractor layer is then computed as $ \bm{x}_{l} = g\left( \bm{W}_l \bm{x}_{l-1} + \bm{b}_l \right), \text{ for } l = 1, \ldots, L$ where $g(\cdot)$ is a non-linear activation function. The final set of extracted feature $\bm{x}_L \in \mathbb{R}^{d_{L,o}}$ is then fed to each of the sub-nets for action predictions for each UE.

Actor sub-net for each UE uses a single layer for base station selection as most of the processing for $\mu BS$ selection from input has been already done by the common feature extractor network. The actor sub-net for $i{th}$ UE predicts a normalized score over all the $\mu BS$ indices using a softmax layer as
    \begin{align}
        \bm{a}^{(i)}_{bs} = \text{softmax}(\bm{W}_{i,bs} \bm{x}_{L} + \bm{b}_{i,bs}),
            \label{eqn:actor_bs_sel}
    \end{align}
    where $\bm{W}_{i,bs} \in \mathbb{R}^{N_{BS} \times d_{L,o}}$ and $\bm{b}_{i,bs} \in \mathbb{R}^{N_{BS}}$. Then, the $\mu BS$ to serve $i^{th}$ UE is selected as base station with highest normalized score in $\bm{a}^{(i)}_{bs}$.
    
    For a better convergence to the best $\mu BS$ for a particular UE, the proposed algorithm is made to explore the space of $\mu BS$ with an $\epsilon$-greedy like mechanism. In the $\epsilon-$greedy method, an agent explores a random discrete action with probability $\epsilon$ and takes an action suggested by the network with probability $(1-\epsilon)$. In the current set-up, the action space is pseudo discrete i.e. the output regarding the selection of $\mu BS$ is a softmax vector (not discrete). These pseudo discrete values are again used to compute the continuous valued angles further. To translate the $\epsilon$-greedy like exploration to our setup and still keep the network differentiable to backpropagate, we propose the following strategy: instead of taking a completely different discrete action, with a probability of $\epsilon_{bs}$, a standard Gaussian noise with high variance is added to the pseudo-discrete normalized score $\bm{a}^{(i)}_{bs}$. However with a probability $(1-\epsilon_{bs})$, the base station with highest normalized score in $\bm{a}^{(i)}_{bs}$ is exploited. Therefore the modified output regarding the $\mu BS$ selection from the actor subnet of each UE is:
    \begin{align}
        \Tilde{\bm{a}}^{(i)}_{bs} = \begin{cases} 
        \bm{a}^{(i)}_{bs} & \text{w.p. } (1-\epsilon_{bs}) \\
        \bm{a}^{(i)}_{bs}+n & \text{w.p. } \epsilon_{bs} \text{ and } n \sim \mathcal{N}(0,1). 
        \end{cases}
        \label{eqn:actor_bs_sel_noisy}
    \end{align}
    The degree of exploration of the algorithm reduces over time i.e. $\epsilon_{bs}$ is an exponentially decaying function with rate $10^{-6}$.

    \begin{algorithm}[!t] 
        \caption{Proposed algorithm for $\mu$BS selection and beam alignment}
        \begin{algorithmic}[1]
            \State \textbf{Parameters:} Set discounting factor $\gamma$, replay buffer size $\tau$,
                    number of episodes $M$, target update period $U$, update parameter $\lambda$, 
                    learning rates $\eta_a$ and $\eta_c$.
            \State Initialize the actor $\mathbbm{A}(s|\bm{\omega}^a)$ and the 
                    critic $\mathbbm{C}(s,a|\bm{\omega}^c)$ networks with random weights
                    $\bm{\omega}^a$ and $\bm{\omega}^c$ respectively.
            \State Initialize target networks with weights as $\bar{\bm{\omega}}^a \gets 
                    \bm{\omega}^a$ and $\bar{\bm{\omega}}^c \gets \bm{\omega}^c$.
            \State Create an empty replay buffer $\mathcal{B} \gets \{\}$ with size $\tau$.
            \For {episode $= 1 \ldots M$}
                \State Select a random valid action for each UE.
                \State Observe SINR as well as the RF signature of each UE as state $s_t$.
                \For {$t = 1 \ldots T$}
                    
                    \State Set $\mu$BS and beam alignment angles for each UE based on the output predicted by the network according to (\ref{eqn:actor_bs_sel_noisy}) and (\ref{eqn:compute_angle_noisy}).
                    \State Get new state observation $s_{t+1}$.
                    \State Extract individual SINR for each UE from $s_{t+1}$ and compute
                            average rate as reward $r_t$.

                    
                    \State Update replay buffer with experience as $\mathcal{B} \gets \mathcal{B} \cup (s_t, a_t, r_t, s_{t+1})$.
                    \If {$|\mathcal{B}| \geq \tau$}
                        \State Delete oldest experience from $\mathcal{B}$.
                    \EndIf
                    \State Sample $N$ experiences $(s_i, a_i, r_i, s_{i+1})$ from $\mathcal{B}$.
                    \State Compute return for each experience $y_i$.
                    \State Compute Mean Square Bellman Error as $\mathcal{L}$.
                    \State Update $(\bm{\omega}^c,\bm{\omega}^a)$ and
                            $(\bar{\bm{\omega}^c}, \bar{\bm{\omega}^a})$.
                \EndFor
            \EndFor
        \end{algorithmic}
        \label{alg:proposed_beamer}
    \end{algorithm}

    Noting that the elevation and the azimuth angles for beam alignment need to depend both on information on the UE positions (available through $\bm{x}_L$) and the selected $\mu BS$ for UE (available through $\bm{a}^{(i)}_{bs}$), we need a layer which can fuse together these information. For this, we first create a concatenating feature vector for each UE as $\bm{z}_{i} = [\bm{x}_L, \bm{a}^{(i)}_{bs}] \in \mathbb{R}^{d_{L,o}+N_{BS}}$. Then, the action corresponding to beam alignment angles for $i^{th}$ UE are computed as
    \begin{align}
        \bm{a}^{(i)}_{\Theta} = \text{tanh}(\bm{W}_{i,\Theta} \bm{z}_i,  + \bm{b}_{i,\Theta}),
        \label{eqn:compute_angle}
    \end{align}
    where $\bm{W}_{i,\Theta} \in \mathbb{R}^{2 \times (d_{L,o}+N_{BS})}$ and $\bm{b}_{i,\Theta} \in \mathbb{R}^2$. Note that the tanh($\cdot$) activation function outputs values in range $[-1, +1]$ and hence $\bm{a}^{(i)}_{\Theta} \in [-1, +1]^2$. To explore the space of elevation and azimuth angles, a Gaussian random normal noise $\mathcal{N}(0,\sigma_{\Theta})$ is added to $\bm{a}^{(i)}_{\Theta}$ where the variance $\sigma_{\Theta}$ decays linearly with time. The modified output from the actor subnet regarding the elevation and azimuth angles is given by:
    \begin{align}
        \Tilde{\bm{a}}^{(i)}_{\Theta} = \bm{a}^{(i)}_{\Theta}+n,
        \label{eqn:compute_angle_noisy}
    \end{align}
    where $n \sim \mathcal{N}(0,\sigma_{\Theta})$. As the overall performance is more sensitive to the selection of elevation and azimuth angles than the best $\mu BS$, a linearly decaying $\sigma_{\Theta}$ (higher exploration) is used for the selection of angles whereas an exponentially decaying $\epsilon_{bs}$ is used (comparatively lesser exploration) for $\mu BS$ selection a discussed before. Finally, the elevation and azimuth angles for beam alignment are computed (in radians) as
    $
        \theta_i = \frac{3\pi}{4} + \bm{a}^{(i)}_{\Theta}[1] \times \frac{\pi}{4},
                    \label{eqn:compute_theta} \text{ and }
        \phi_i = \bm{a}^{(i)}_{\Theta}[2] \times \frac{\pi}{2}.
                    \label{eqn:compute_phi}
    $
    The computation for elevation angle is based on the assumption that all UEs are below the height of $\mu BS$ and hence $\theta_i \in [\pi/2, \pi]$. Similarly, it is assumed that $\phi_i \in 
    [-\pi/2, +\pi/2]$.

The proposed algorithm to train the DRL agent for base station selection 
and beam alignment is presented in Alg. \ref{alg:proposed_beamer}.

To appreciate the power of the proposed method, we also perform two experiments named \textit{BSOracle} and \textit{AngleOracle} which serve as benchmarks. In the first method, the agent has the exact knowledge of the $\mu BS$ and in the latter, the location of the UEs and hence the elevation and the azimuth angles are known exactly by the agent. The main differences between the proposed method and BSOracle/AngleOracle methods are listed below.

\subsubsection{\textbf{Architecture of Neural action predictor of BSOracle method}}
The proposed method takes $s_t\in\mathbb{R}^{N_{UE}.(N_{BS}+1)}$ as an input at time $t$ to the learning agent. The input to the proposed method has only the RF signatures from $N_{BS}$ $\mu BS$s and the SINR for each $N_{UE}$ UEs. Unlike the proposed method, the input to the learning agent in the case of BSOracle is a $N_{UE}.(N_{BS}+N_{BS}+1)$ dimensional vector (observations of $N_{UE}$ UEs are concatenated) which is fed to the common feature extractor network as input. For each UE, the first $N_{BS}$ elements are the one-hot encoded vector indicating the best $\mu BS$ for each UE. The next $N_{BS}$ elements are RF fingerprints and the last element is the SINR of that UE for the current beam alignment configuration. The output of the $l^{th}$ feature extractor layer is computed as
$
    \bm{x}_{l} = g\left( \bm{W}_l \bm{x}_{l-1} + \bm{b}_l \right),
            \text{ for } l = 1, \ldots, L
$
where $g(.)$ is the non-linear activation function. The output of the feature network $\bm{x}_{L}$ is fed to each of the UE sub-nets those are used to predict \textit{just the elevation and azimuth angles} because the best $\mu BS$ is already known by the agent via the input features. The action corresponding to the beam alignment angles for the $i^{th}$ UE are computed as 
\begin{align}
    \bm{a}^{(i)}_{\Theta} = \text{tanh}(\bm{W}_{i,\Theta} \bm{x}_{L},  + \bm{b}_{i,\Theta}),
        \label{eqn:compute_angle_BSOracle}
\end{align}
where $\bm{W}_{i,\Theta} \in \mathbb{R}^{2 \times d_{L,o}}$ and $\bm{b}_{i,\Theta} \in \mathbb{R}^2$. Similar to the proposed method, a Gaussian noise with decaying variance is used to explore the space of angles in a better way. Finally the angles are converted to radians following the same way like the proposed method discussed before. The final action is the concatenation of the vector referring the index of the best $\mu BS$ for each UE and the elevation and azimuth angles. 

\subsubsection{\textbf{Architecture of Neural action predictor of AngleOracle method}}
Similar to the proposed method, the input $s_t$ to the learning agent at time $t$ is a $N_{UE}.(N_{BS}+1)$ dimensional vector. The rest of the architecture is very similar to the proposed method except that each UE sub-net predicts the pseudo discrete output that represents the index of the best $\mu BS$. Using this information the genie predicts the correct elevation and azimuth angles.

We reiterate that both AngleOracle and BSOracle only serve as possible benchmarks against which we can compare the proposed approach and these by themselves cannot be implemented in a practical scenario since one will never know either the best $\mu BS$ for each UE or the UE locations.

    \section{Results}

\ifCLASSOPTIONtwocolumn
    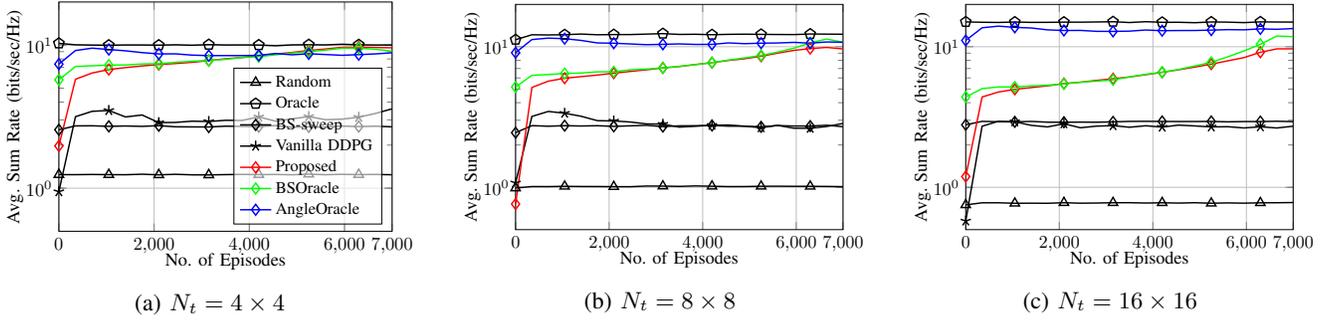
\begin{figure*}[!t]
        \centering
        \begin{subfigure}{.33\linewidth}
            \resizebox{\linewidth}{!}{
                \pgfplotstableread[col sep = comma]{./data/00_Nancy/data_03UE_BS04x04_rateEvo_Baselines_NANZ.csv}\datatable
                \pgfplotstableread[col sep = comma]{./data/00_Nancy/data_03UE_BS04x04_rateEvoNANZ-BS-noisyaction21.csv}\datatablenancysigmaone 
                \pgfplotstableread[col sep = comma]{./data/00_Nancy/data_03UE_BS04x04_rateEvoNANZ-BSKnown.csv}\datatablenancybsknownrateevo
               \pgfplotstableread[col sep = comma]{./data/00_Nancy/data_03UE_BS04x04_rateEvoNANZ-AngleKnown.csv}\datatablenancyangleknownrateevo
            \tikzstyle{mark_style} = []

\begin{tikzpicture}[thick,scale=0.8]
    \begin{semilogyaxis}[
        width=8cm,
        height=6cm,
        xmin=0,
        xmax=7000,
        ymin=0.5,
        ymax=20,
        grid=major,
        xlabel={No. of Episodes},
        ylabel={Avg. Sum Rate (bits/sec/Hz)},
        xlabel style={at={(0.50,0.05)}},
        ylabel style={at={(0.06,0.50)}},
        ytick={0,1,10},
        xtick={0,2000, 4000, 6000, 7000},
        legend pos=south east,
        legend cell align={left},
        legend style={fill opacity=0.6, draw opacity=1.0, text opacity=1.0, font=\small}
        ]
        
        \addplot[black, solid, thick, mark=triangle, mark size={3.0}, mark_style, mark repeat=3,
                ] 
            table [x=x_data, y expr=\thisrow{y_rand}/1000, col sep=comma]{\datatable};
        \addlegendentry{Random};

        \addplot[black, solid, thick, mark=pentagon, mark size={3.0}, mark_style, mark repeat=3,
                ] 
            table [x=x_data, y expr=\thisrow{y_greedy}/1000, col sep=comma]{\datatable};
        \addlegendentry{Oracle};

        \addplot[black, solid, thick, 
                mark=diamond, mark size={3.0}, mark size={3.0}, mark_style, mark repeat=3,
                ] 
            table [x=x_data, y expr=\thisrow{y_sweep24}/1000, col sep=comma]{\datatable};
        \addlegendentry{BS-sweep};

        \addplot[black, solid, thick, 
                mark=star, mark size={3.0}, mark size={3.0}, mark_style, mark repeat=3,
                ] 
            table [x=x_data, y expr=\thisrow{y_ddpg_df_2x128_gamma060}/1000, col sep=comma]{\datatable};
        \addlegendentry{Vanilla DDPG};

        \addplot[red, solid, thick, 
                mark=diamond, mark size={3.0}, mark size={3.0}, mark_style, mark repeat=3,
                ] 
            table [x=x_data, y expr=\thisrow{y_ddpg_2x128_gamma060}/1000, col sep=comma]{\datatablenancysigmaone};
        \addlegendentry{Proposed};
        
        \addplot[green, solid, thick, 
                mark=diamond, mark size={3.0}, mark size={3.0}, mark_style, mark repeat=3,
                ] 
            table [x=x_data, y expr=\thisrow{y_ddpg_2x128_gamma060}/1000, col sep=comma]{\datatablenancybsknownrateevo};
        \addlegendentry{BSOracle};
        
        \addplot[blue, solid, thick, 
                mark=diamond, mark size={3.0}, mark size={3.0}, mark_style, mark repeat=3,
                ] 
            table [x=x_data, y expr=\thisrow{y_ddpg_2x128_gamma060}/1000, col sep=comma]{\datatablenancyangleknownrateevo};
        \addlegendentry{AngleOracle};
    \end{semilogyaxis}
\end{tikzpicture}
            }
            \caption{$N_t = 4 \times 4$}
            \label{fig:rateevo_03ue_bs04x04}
        \end{subfigure}%
        \begin{subfigure}{.33\linewidth}
            \resizebox{\linewidth}{!}{
                \pgfplotstableread[col sep = comma]{./data/00_Nancy/data_03UE_BS08x08_rateEvo_Baselines_NANZ.csv}\datatable
                \pgfplotstableread[col sep = comma]{./data/00_Nancy/data_03UE_BS08x08_rateEvoNANZ-BS-noisyaction21.csv}\datatablenancysigmaone
                \pgfplotstableread[col sep = comma]{./data/00_Nancy/data_03UE_BS08x08_rateEvoNANZ-BSKnown.csv}\datatablenancybsknownrateevo
                \pgfplotstableread[col sep = comma]{./data/00_Nancy/data_03UE_BS08x08_rateEvoNANZ-AngleKnown.csv}\datatablenancyangleknownrateevo
                \pgfplotsset{ignore legend}
                \tikzstyle{mark_style} = []
\begin{tikzpicture}[thick,scale=0.8]
    \begin{semilogyaxis}[
        width=8cm,
        height=6cm,
        xmin=0,
        xmax=7000,
        ymin=0.5,
        ymax=20,
        grid=major,
        xlabel={No. of Episodes},
        ylabel={Avg. Sum Rate (bits/sec/Hz)},
        xlabel style={at={(0.50,0.05)}},
        ylabel style={at={(0.06,0.50)}},
        ytick={0,1,10},
        xtick={0,2000, 4000, 6000, 7000},
        legend pos=south east,
        legend cell align={left},
        legend style={fill opacity=0.6, draw opacity=1.0, text opacity=1.0, font=\small}
        ]
        
        \addplot[black, solid, thick, mark=triangle, mark size={3.0}, mark_style, mark repeat=3,
                ] 
            table [x=x_data, y expr=\thisrow{y_rand}/1000, col sep=comma]{\datatable};
        \addlegendentry{Random};

        \addplot[black, solid, thick, mark=pentagon, mark size={3.0}, mark_style, mark repeat=3,
                ] 
            table [x=x_data, y expr=\thisrow{y_greedy}/1000, col sep=comma]{\datatable};
        \addlegendentry{Oracle};

        \addplot[black, solid, thick, 
                mark=diamond, mark size={3.0}, mark size={3.0}, mark_style, mark repeat=3,
                ] 
            table [x=x_data, y expr=\thisrow{y_sweep24}/1000, col sep=comma]{\datatable};
        \addlegendentry{BS-sweep};

        \addplot[black, solid, thick, 
                mark=star, mark size={3.0}, mark size={3.0}, mark_style, mark repeat=3,
                ] 
            table [x=x_data, y expr=\thisrow{y_ddpg_df_2x128_gamma060}/1000, col sep=comma]{\datatable};
        \addlegendentry{Vanilla DDPG};

        \addplot[red, solid, thick, 
                mark=diamond, mark size={3.0}, mark size={3.0}, mark_style, mark repeat=3,
                ] 
            table [x=x_data, y expr=\thisrow{y_ddpg_2x128_gamma060}/1000, col sep=comma]{\datatablenancysigmaone};
        \addlegendentry{Proposed};
        
        \addplot[green, solid, thick, 
                mark=diamond, mark size={3.0}, mark size={3.0}, mark_style, mark repeat=3,
                ] 
            table [x=x_data, y expr=\thisrow{y_ddpg_2x128_gamma060}/1000, col sep=comma]{\datatablenancybsknownrateevo};
        \addlegendentry{BSOracle};
        
        \addplot[blue, solid, thick, 
                mark=diamond, mark size={3.0}, mark size={3.0}, mark_style, mark repeat=3,
                ] 
            table [x=x_data, y expr=\thisrow{y_ddpg_2x128_gamma060}/1000, col sep=comma]{\datatablenancyangleknownrateevo};
        \addlegendentry{AngleOracle};
    \end{semilogyaxis}
\end{tikzpicture}
            }
            \caption{$N_t = 8 \times 8$}
            \label{fig:rateevo_03ue_bs08x08}
        \end{subfigure}%
        \begin{subfigure}{.33\linewidth}
            \resizebox{\linewidth}{!}{
                \pgfplotstableread[col sep = comma]{./data/00_Nancy/data_03UE_BS16x16_rateEvo_Baselines_NANZ.csv}\datatable
                \pgfplotstableread[col sep = comma]{./data/02_psuedo_act/data_03UE_BS16x16_rateEvov1.csv}\tabpsuedoactvone
                \pgfplotstableread[col sep = comma]{./data/00_Nancy/data_03UE_BS16x16_rateEvoNANZ-BS-noisyaction21.csv}\datatablenancysigmaone
                \pgfplotstableread[col sep = comma]{./data/00_Nancy/data_03UE_BS16x16_rateEvoNANZ-BSKnown.csv}\datatablenancybsknownrateevo
                \pgfplotstableread[col sep = comma]{./data/00_Nancy/data_03UE_BS16x16_rateEvoNANZ-AngleKnown.csv}\datatablenancyangleknownrateevo
                \pgfplotsset{ignore legend}
                \tikzstyle{mark_style} = []
\begin{tikzpicture}[thick,scale=0.8]
    \begin{semilogyaxis}[
        width=8cm,
        height=6cm,
        xmin=0,
        xmax=7000,
        ymin=0.5,
        ymax=20,
        grid=major,
        xlabel={No. of Episodes},
        ylabel={Avg. Sum Rate (bits/sec/Hz)},
        xlabel style={at={(0.50,0.05)}},
        ylabel style={at={(0.06,0.50)}},
        ytick={0,1,10},
        xtick={0,2000, 4000, 6000, 7000},
        legend pos=south east,
        legend cell align={left},
        legend style={fill opacity=0.6, draw opacity=1.0, text opacity=1.0, font=\small}
        ]
        
        \addplot[black, solid, thick, mark=triangle, mark size={3.0}, mark_style, mark repeat=3,
                ] 
            table [x=x_data, y expr=\thisrow{y_rand}/1000, col sep=comma]{\datatable};
        \addlegendentry{Random};

        \addplot[black, solid, thick, mark=pentagon, mark size={3.0}, mark_style, mark repeat=3,
                ] 
            table [x=x_data, y expr=\thisrow{y_greedy}/1000, col sep=comma]{\datatable};
        \addlegendentry{Oracle};

        \addplot[black, solid, thick, 
                mark=diamond, mark size={3.0}, mark size={3.0}, mark_style, mark repeat=3,
                ] 
            table [x=x_data, y expr=\thisrow{y_sweep24}/1000, col sep=comma]{\datatable};
        \addlegendentry{BS-sweep};

        \addplot[black, solid, thick, 
                mark=star, mark size={3.0}, mark size={3.0}, mark_style, mark repeat=3,
                ] 
            table [x=x_data, y expr=\thisrow{y_ddpg_df_2x128_gamma060}/1000, col sep=comma]{\datatable};
        \addlegendentry{Vanilla DDPG};

        \addplot[red, solid, thick, 
                mark=diamond, mark size={3.0}, mark size={3.0}, mark_style, mark repeat=3,
                ] 
            table [x=x_data, y expr=\thisrow{y_ddpg_2x128_gamma060}/1000, col sep=comma]{\datatablenancysigmaone};
        \addlegendentry{Proposed};
        
        \addplot[green, solid, thick, 
                mark=diamond, mark size={3.0}, mark size={3.0}, mark_style, mark repeat=3,
                ] 
            table [x=x_data, y expr=\thisrow{y_ddpg_2x128_gamma060}/1000, col sep=comma]{\datatablenancybsknownrateevo};
        \addlegendentry{BSOracle};
        
        \addplot[blue, solid, thick, 
                mark=diamond, mark size={3.0}, mark size={3.0}, mark_style, mark repeat=3,
                ] 
            table [x=x_data, y expr=\thisrow{y_ddpg_2x128_gamma060}/1000, col sep=comma]{\datatablenancyangleknownrateevo};
        \addlegendentry{AngleOracle};
    \end{semilogyaxis}
\end{tikzpicture}
            }
            \caption{$N_t = 16 \times 16$}
            \label{fig:rateevo_03ue_bs16x16}
        \end{subfigure}%
        \caption{Rate Evolution during learning for $N_{UE} = 3$.}
        \label{fig:rateevo_03ue}
    \end{figure*}
    
    \begin{figure*}[!t]
        \centering
        \begin{subfigure}{.33\linewidth}
            \resizebox{\linewidth}{!}{
                \pgfplotstableread[col sep = comma]{./data/00_Nancy/data_05UE_BS04x04_rateEvo_Baselines_NANZ.csv}\datatable
              \pgfplotstableread[col sep = comma]{./data/00_Nancy/data_05UE_BS04x04_rateEvoNANZ-BS-noisyaction21.csv}\datatablenancysigmaone 
              \pgfplotstableread[col sep = comma]{./data/00_Nancy/data_05UE_BS04x04_rateEvoNANZ-BSKnown.csv}\datatablenancybsknownrateevo
              \pgfplotstableread[col sep = comma]{./data/00_Nancy/data_05UE_BS04x04_rateEvoNANZ-AngleKnown.csv}\datatablenancyangleknownrateevo
              \pgfplotsset{ignore legend}
            \tikzstyle{mark_style} = []
\begin{tikzpicture}[thick,scale=0.8]
    \begin{semilogyaxis}[
        width=8cm,
        height=6cm,
        xmin=0,
        xmax=7000,
        ymin=0.5,
        ymax=20,
        grid=major,
        xlabel={No. of Episodes},
        ylabel={Avg. Sum Rate (bits/sec/Hz)},
        xlabel style={at={(0.50,0.05)}},
        ylabel style={at={(0.06,0.50)}},
        ytick={0,1,10},
        xtick={0,2000, 4000, 6000, 7000},
        legend pos=south east,
        legend cell align={left},
        legend style={fill opacity=0.6, draw opacity=1.0, text opacity=1.0, font=\small}
        ]
        
        \addplot[black, solid, thick, mark=triangle, mark size={3.0}, mark_style, mark repeat=3,
                ] 
            table [x=x_data, y expr=\thisrow{y_rand}/1000, col sep=comma]{\datatable};
        \addlegendentry{Random};

        \addplot[black, solid, thick, mark=pentagon, mark size={3.0}, mark_style, mark repeat=3,
                ] 
            table [x=x_data, y expr=\thisrow{y_greedy}/1000, col sep=comma]{\datatable};
        \addlegendentry{Oracle};

        \addplot[black, solid, thick, 
                mark=diamond, mark size={3.0}, mark size={3.0}, mark_style, mark repeat=3,
                ] 
            table [x=x_data, y expr=\thisrow{y_sweep24}/1000, col sep=comma]{\datatable};
        \addlegendentry{BS-sweep};

        \addplot[black, solid, thick, 
                mark=star, mark size={3.0}, mark size={3.0}, mark_style, mark repeat=3,
                ] 
            table [x=x_data, y expr=\thisrow{y_ddpg_df_2x128_gamma060}/1000, col sep=comma]{\datatable};
        \addlegendentry{Vanilla DDPG};

        \addplot[red, solid, thick, 
                mark=diamond, mark size={3.0}, mark size={3.0}, mark_style, mark repeat=3,
                ] 
            table [x=x_data, y expr=\thisrow{y_ddpg_2x128_gamma060}/1000, col sep=comma]{\datatablenancysigmaone};
        \addlegendentry{Proposed};
        
        \addplot[green, solid, thick, 
                mark=diamond, mark size={3.0}, mark size={3.0}, mark_style, mark repeat=3,
                ] 
            table [x=x_data, y expr=\thisrow{y_ddpg_2x128_gamma060}/1000, col sep=comma]{\datatablenancybsknownrateevo};
        \addlegendentry{BSOracle};
        
        \addplot[blue, solid, thick, 
                mark=diamond, mark size={3.0}, mark size={3.0}, mark_style, mark repeat=3,
                ] 
            table [x=x_data, y expr=\thisrow{y_ddpg_2x128_gamma060}/1000, col sep=comma]{\datatablenancyangleknownrateevo};
        \addlegendentry{AngleOracle};
    \end{semilogyaxis}
\end{tikzpicture}
            }
            \caption{$N_t = 4 \times 4$}
            \label{fig:rateevo_05ue_bs04x04}
        \end{subfigure}%
        \begin{subfigure}{.33\linewidth}
            \resizebox{\linewidth}{!}{
                \pgfplotstableread[col sep = comma]{./data/00_Nancy/data_05UE_BS08x08_rateEvo_Baselines_NANZ.csv}\datatable
                \pgfplotstableread[col sep = comma]{./data/02_psuedo_act/data_05UE_BS08x08_rateEvov1.csv}\tabpsuedoactvone
                \pgfplotstableread[col sep = comma]{./data/00_Nancy/data_05UE_BS08x08_rateEvoNANZ-BS-noisyaction21.csv}\datatablenancysigmaone
                \pgfplotstableread[col sep = comma]{./data/00_Nancy/data_05UE_BS08x08_rateEvoNANZ-BSKnown.csv}\datatablenancybsknownrateevo
                \pgfplotstableread[col sep = comma]{./data/00_Nancy/data_05UE_BS08x08_rateEvoNANZ-AngleKnown.csv}\datatablenancyangleknownrateevo
                \pgfplotsset{ignore legend}
                \tikzstyle{mark_style} = []
\begin{tikzpicture}[thick,scale=0.8]
    \begin{semilogyaxis}[
        width=8cm,
        height=6cm,
        xmin=0,
        xmax=7000,
        ymin=0.5,
        ymax=20,
        grid=major,
        xlabel={No. of Episodes},
        ylabel={Avg. Sum Rate (bits/sec/Hz)},
        xlabel style={at={(0.50,0.05)}},
        ylabel style={at={(0.06,0.50)}},
        ytick={0,1,10},
        xtick={0,2000, 4000, 6000, 7000},
        legend pos=south east,
        legend cell align={left},
        legend style={fill opacity=0.6, draw opacity=1.0, text opacity=1.0, font=\small}
        ]
        
        \addplot[black, solid, thick, mark=triangle, mark size={3.0}, mark_style, mark repeat=3,
                ] 
            table [x=x_data, y expr=\thisrow{y_rand}/1000, col sep=comma]{\datatable};
        \addlegendentry{Random};

        \addplot[black, solid, thick, mark=pentagon, mark size={3.0}, mark_style, mark repeat=3,
                ] 
            table [x=x_data, y expr=\thisrow{y_greedy}/1000, col sep=comma]{\datatable};
        \addlegendentry{Oracle};

        \addplot[black, solid, thick, 
                mark=diamond, mark size={3.0}, mark size={3.0}, mark_style, mark repeat=3,
                ] 
            table [x=x_data, y expr=\thisrow{y_sweep24}/1000, col sep=comma]{\datatable};
        \addlegendentry{BS-sweep};

        \addplot[black, solid, thick, 
                mark=star, mark size={3.0}, mark size={3.0}, mark_style, mark repeat=3,
                ] 
            table [x=x_data, y expr=\thisrow{y_ddpg_df_2x128_gamma060}/1000, col sep=comma]{\datatable};
        \addlegendentry{Vanilla DDPG};

        \addplot[red, solid, thick, 
                mark=diamond, mark size={3.0}, mark size={3.0}, mark_style, mark repeat=3,
                ] 
            table [x=x_data, y expr=\thisrow{y_ddpg_2x128_gamma060}/1000, col sep=comma]{\datatablenancysigmaone};
        \addlegendentry{Proposed};
        
        \addplot[green, solid, thick, 
                mark=diamond, mark size={3.0}, mark size={3.0}, mark_style, mark repeat=3,
                ] 
            table [x=x_data, y expr=\thisrow{y_ddpg_2x128_gamma060}/1000, col sep=comma]{\datatablenancybsknownrateevo};
        \addlegendentry{BSOracle};
        
        \addplot[blue, solid, thick, 
                mark=diamond, mark size={3.0}, mark size={3.0}, mark_style, mark repeat=3,
                ] 
            table [x=x_data, y expr=\thisrow{y_ddpg_2x128_gamma060}/1000, col sep=comma]{\datatablenancyangleknownrateevo};
        \addlegendentry{AngleOracle};
    \end{semilogyaxis}
\end{tikzpicture}
            }
            \caption{$N_t = 8 \times 8$}
            \label{fig:rateevo_05ue_bs08x08}
        \end{subfigure}%
        \begin{subfigure}{.33\linewidth}
            \resizebox{\linewidth}{!}{
                \pgfplotstableread[col sep = comma]{./data/00_Nancy/data_05UE_BS16x16_rateEvo_Baselines_NANZ.csv}\datatable
                
                \pgfplotstableread[col sep = comma]{./data/00_Nancy/data_05UE_BS16x16_rateEvoNANZ-BS-noisyaction21.csv}\datatablenancysigmaone
                
                \pgfplotstableread[col sep = comma]{./data/00_Nancy/data_05UE_BS16x16_rateEvoNANZ-BSKnown.csv}\datatablenancybsknownrateevo
                \pgfplotstableread[col sep = comma]{./data/00_Nancy/data_05UE_BS16x16_rateEvoNANZ-AngleKnown.csv}\datatablenancyangleknownrateevo
                \pgfplotsset{ignore legend}
                \tikzstyle{mark_style} = []
\begin{tikzpicture}[thick,scale=0.8]
    \begin{semilogyaxis}[
        width=8cm,
        height=6cm,
        xmin=0,
        xmax=7000,
        ymin=0.5,
        ymax=20,
        grid=major,
        xlabel={No. of Episodes},
        ylabel={Avg. Sum Rate (bits/sec/Hz)},
        xlabel style={at={(0.50,0.05)}},
        ylabel style={at={(0.06,0.50)}},
        ytick={0,1,10},
        xtick={0,2000, 4000, 6000, 7000},
        legend pos=south east,
        legend cell align={left},
        legend style={fill opacity=0.6, draw opacity=1.0, text opacity=1.0, font=\small}
        ]
        
        \addplot[black, solid, thick, mark=triangle, mark size={3.0}, mark_style, mark repeat=3,
                ] 
            table [x=x_data, y expr=\thisrow{y_rand}/1000, col sep=comma]{\datatable};
        \addlegendentry{Random};

        \addplot[black, solid, thick, mark=pentagon, mark size={3.0}, mark_style, mark repeat=3,
                ] 
            table [x=x_data, y expr=\thisrow{y_greedy}/1000, col sep=comma]{\datatable};
        \addlegendentry{Oracle};

        \addplot[black, solid, thick, 
                mark=diamond, mark size={3.0}, mark size={3.0}, mark_style, mark repeat=3,
                ] 
            table [x=x_data, y expr=\thisrow{y_sweep24}/1000, col sep=comma]{\datatable};
        \addlegendentry{BS-sweep};

        \addplot[black, solid, thick, 
                mark=star, mark size={3.0}, mark size={3.0}, mark_style, mark repeat=3,
                ] 
            table [x=x_data, y expr=\thisrow{y_ddpg_df_2x128_gamma060}/1000, col sep=comma]{\datatable};
        \addlegendentry{Vanilla DDPG};

        \addplot[red, solid, thick, 
                mark=diamond, mark size={3.0}, mark size={3.0}, mark_style, mark repeat=3,
                ] 
            table [x=x_data, y expr=\thisrow{y_ddpg_2x128_gamma060}/1000, col sep=comma]{\datatablenancysigmaone};
        \addlegendentry{Proposed};
        
        \addplot[green, solid, thick, 
                mark=diamond, mark size={3.0}, mark size={3.0}, mark_style, mark repeat=3,
                ] 
            table [x=x_data, y expr=\thisrow{y_ddpg_2x128_gamma060}/1000, col sep=comma]{\datatablenancybsknownrateevo};
        \addlegendentry{BSOracle};
        
        \addplot[blue, solid, thick, 
                mark=diamond, mark size={3.0}, mark size={3.0}, mark_style, mark repeat=3,
                ] 
            table [x=x_data, y expr=\thisrow{y_ddpg_2x128_gamma060}/1000, col sep=comma]{\datatablenancyangleknownrateevo};
        \addlegendentry{AngleOracle};
    \end{semilogyaxis}
\end{tikzpicture}
            }
            \caption{$N_t = 16 \times 16$}
            \label{fig:rateevo_05ue_bs16x16}
        \end{subfigure}%
        \caption{Rate Evolution during learning for $N_{UE} = 5$.}
        \label{fig:rateevo_05ue}
    \end{figure*}
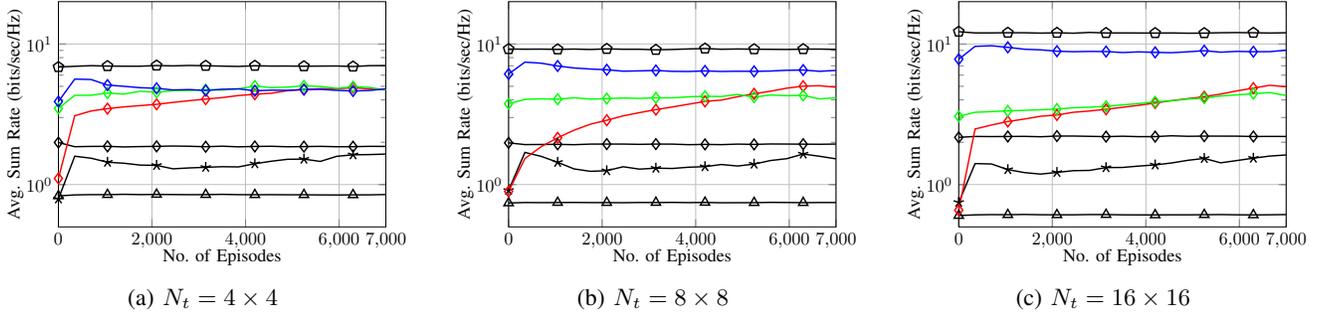
\fi

In this section, we provide simulation results for a four-junction scenario similar to \cite{alkhateeb2018deep} (extended to four roads) with $N_{BS} = 10$ and an intercell radius of approximately $100m$. A carrier frequency of $f_c = 28GHz$ is assumed and bandwidth of $5MHz$ is taken. All $\mu BS$ are assumed to have UPA antenna of square dimensions with $N_t = 4 \times 4$ and with $d = \lambda/2$, where is $\lambda$ is the wavelength associated with frequency $f_c$. Following the Street Canyon configuration \cite{5gmodel2016}, we used $n = 1.98$, $\sigma=3.1$, $b = 0$ and $f_0 = 10^{9}$ as the path loss parameters in (\ref{eqn:pathloss}). We provide results for $N_{UE} = 3, \text{ and } 5$ with $\kappa = 1$.

The performance of the proposed method and the baseline competitor methods are compared below:
\begin{enumerate}
    \item \textbf{Random}: A blind agent that does not receive any inputs about the UEs, but tries to assign a $\mu BS$ and a set of beam alignment angles for each UE. As this algorithm does not have any input/feedback, the rate obtained by this method is the minimum expected rate that can be obtained by any \textit{intelligent} agent. \item \textbf{Oracle}: This agent assumes that the exact knowledge about the location of UEs as well as the exact channel is known at the BS. Equipped with this information, \textit{Oracle} picks the best $\mu$BS-UE assignment as well as the beam alignment angles. This is the maximum expected rate any algorithm can achieve.
    \item \textbf{BS-Sweep}: This method is similar to the one proposed in \cite{nitsche2014ieee}. However, as there are multiple $\mu$BS, all BS will simultaneously perform brute force beam search with a beam at every $\nu$ degrees. Since some of the available time is spent on the UE discovery process, the reported metrics are based on the available transmit time. Also, UE discovery needs to be performed at every instant as the UEs are mobile. We considered a frame period of $10ms$ and a beam scan period of $200\mu s$ per beam. While more number of beams increases the resolution of UE discovery, it also adds an overhead time. We provide results for $\nu = 5 \deg$.
    \item \textbf{Vanilla DDPG}: This is an RL agent that uses the feedforward neural network as proposed in \cite{lillicrap2015continuous} in the context of game-playing agents. This is provided to quantify the improvement in the performance of the proposed neural network architecture. The neural network considered has $L = 2$. Each hidden layer has $128$ nodes. The difference between this and the proposed method is that the UE sub-net is absent in this method. We provide the results averaged over $5$ agents, each trained for $7000$ episodes. Each episode is considered $1000$ timesteps long and the UE positions are reset at the end of each episode.
    \item \textbf{Proposed}: The proposed Deep Reinforcement Learning based agent has a feed-forward critic network and UE sub-net augmented actor-network, with $L = 2$. Each hidden layer has $128$ hidden nodes. All other conditions including the training environment and the optimizer parameters are the same as \textit{Vanilla DDPG} mentioned above.
    \item \textbf{BSOracle}: This agent predicts the elevation and azimuth angles with the knowledge about the best $\mu BS$ for each UE. The architecture of this is the same as the proposed method except that each sub-net has only a single hidden layer that predicts the angles. 
    \item \textbf{AngleOracle}: Here the agent predicts only the index of the best $\mu BS$ for each UE and the genie gives the elevation and azimuth angles beased on this. The architecture is same as the proposed method, except that in each sub-net there is only a single layer that predicts the index of the best $\mu BS$. 
\end{enumerate}
In BSOracle method, the best $\mu BS$ for each UE is given by the genie based on which the angles are predicted; whereas in AngleOracle method, the agent predicts only the best $\mu BS$ but the best elevation and azimuth angles for each UE are evaluated by the genie. Needless to say, both BSOracle and AngleOracle methods perform better than the proposed. However, these two methods are carried out to assess the performance of the proposed method that finds the best $\mu BS$ as well as the elevation and azimuth angles for each UE completely by learning.
\begin{table}[!h]
    \centering
    \caption{Parameter for learning algorithms}   \label{tab:drl_params}
    \begin{tabular}{|l|c|}
    \hline
    \textbf{Parameter}      & \textbf{Value} \\
    \hline
    \hline
    Number of Hidden Layers, $L$     & $2$ \\
    \hline
    Hidden Nodes in layer            & $[ 128, 128]$ \\
    \hline
    Buffer Size, $\tau$             & $100000$ \\
    \hline
    Discounting factor, $\gamma$    & $0.60$ \\
    \hline
    $\lambda$                       & $0.001$ \\
    \hline
    Actor learning rate, $\eta_a$   & $0.0001$ \\
    \hline
    Critic learning rate, $\eta_c$  & $0.001$ \\
    \hline
    Number of episodes              & $1000$ \\
    \hline
    Steps per episode               & $1000$ \\
    \hline
    \end{tabular}
\end{table}

The parameters used for training deep reinforcement learning agents are given in Table \ref{tab:drl_params}. The values are found after sufficient hyperparameter tuning through manual search. For a fair comparison, we keep all the parameters the same across Vanilla DDPG, BSOracle, AngleOracle, and the Proposed architecture.

\ifCLASSOPTIONtwocolumn
\begin{figure*}[!t]
    \centering
    \begin{subfigure}{.33\linewidth}
        \resizebox{\linewidth}{!}{
            \pgfplotstableread[col sep = comma]{./data/00_Nancy/data_03UE_BS04x04_rateCdf_Baselines_NANZ.csv}\datatable
         
            \pgfplotstableread[col sep = comma]{./data/00_Nancy/data_03UE_BS04x04_rateCdfNANZ-BS-noisyaction21.csv}\datatablenancyvarsigma
            \pgfplotstableread[col sep = comma]{./data/00_Nancy/data_03UE_BS04x04_rateCdfNANZ-BSKnown.csv}\datatablenancybsknownratecdf
            \pgfplotstableread[col sep = comma]{./data/00_Nancy/data_03UE_BS04x04_rateCdfNANZ-AngleKnown.csv}\datatablenancyangleknownratecdf
            \tikzstyle{mark_style} = [mark size={3.0}, mark repeat=20, mark phase=1]
\begin{tikzpicture}[thick,scale=0.8]
    \begin{axis}[
        width=8cm,
        height=6cm,
        xmin=-1.0,
        xmax=25,
        ymin=-0.1,
        ymax=+1.1,
        grid=major,
        xlabel={Rate (bits/sec/Hz)},
        ylabel={CDF},
        xlabel style={at={(0.50,0.05)}},
        ylabel style={at={(0.06,0.50)}},
        ytick={0.0,0.2,...,1.0},
        legend pos=south east,
        legend cell align={left},
        legend style={fill opacity=0.6, draw opacity=1.0, text opacity=1.0, font=\small}
        ]
        
        \addplot[black, solid, thick, mark=triangle, mark size={3.0}, mark_style
                ] 
            table [y=cdf, x=rand, col sep=comma]{\datatable};
        \addlegendentry{Random};

        \addplot[black, solid, thick, mark=pentagon, mark size={3.0}, mark_style
                ] 
            table [y=cdf, x=greedy, col sep=comma]{\datatable};
        \addlegendentry{Oracle};

        \addplot[black, solid, thick, 
                mark=diamond, mark size={3.0}, mark size={3.0}, mark_style
                ] 
            table [y=cdf, x=sweep24, col sep=comma]{\datatable};
        \addlegendentry{BS-sweep};

        \addplot[black, solid, thick, 
                mark=star, mark size={3.0}, mark size={3.0}, mark_style
                ] 
            table [y=cdf, x=ddpg_df_2x128_gamma060_000, col sep=comma]{\datatable};
        \addlegendentry{Vanilla DDPG};

        \addplot[red, solid, thick, 
                mark=diamond, mark size={3.0}, mark size={3.0}, mark_style
                ] 
            table [y=cdf, x=ddpg_2x128_gamma060_000, col sep=comma]{\datatablenancyvarsigma};
        \addlegendentry{Proposed};
        
        \addplot[green, solid, thick, 
                mark=diamond, mark size={3.0}, mark size={3.0}, mark_style
                ] 
            table [y=cdf, x=ddpg_2x128_gamma060_000, col sep=comma]{\datatablenancybsknownratecdf};
        \addlegendentry{BSOracle};
        
        \addplot[blue, solid, thick, 
                mark=diamond, mark size={3.0}, mark size={3.0}, mark_style
                ] 
            table [y=cdf, x=ddpg_2x128_gamma060_000, col sep=comma]{\datatablenancyangleknownratecdf};
        \addlegendentry{AngleOracle};
    \end{axis}
\end{tikzpicture}
        }
        \caption{$N_t = 4 \times 4$}
        \label{fig:ratecdf_03ue_bs04x04}
    \end{subfigure}%
    \begin{subfigure}{.33\linewidth}
        \resizebox{\linewidth}{!}{
            \pgfplotstableread[col sep = comma]{./data/00_Nancy/data_03UE_BS08x08_rateCdf_Baselines_NANZ.csv}\datatable
            \pgfplotstableread[col sep = comma]{./data/00_Nancy/data_03UE_BS08x08_rateCdfNANZ-BS-noisyaction21.csv}\datatablenancyvarsigma
            \pgfplotstableread[col sep = comma]{./data/00_Nancy/data_03UE_BS08x08_rateCdfNANZ-BSKnown.csv}\datatablenancybsknownratecdf
            \pgfplotstableread[col sep = comma]{./data/00_Nancy/data_03UE_BS08x08_rateCdfNANZ-AngleKnown.csv}\datatablenancyangleknownratecdf
            \pgfplotsset{ignore legend}
            \tikzstyle{mark_style} = [mark size={3.0}, mark repeat=20, mark phase=1]
\begin{tikzpicture}[thick,scale=0.8]
    \begin{axis}[
        width=8cm,
        height=6cm,
        xmin=-1.0,
        xmax=25,
        ymin=-0.1,
        ymax=+1.1,
        grid=major,
        xlabel={Rate (bits/sec/Hz)},
        ylabel={CDF},
        xlabel style={at={(0.50,0.05)}},
        ylabel style={at={(0.06,0.50)}},
        ytick={0.0,0.2,...,1.0},
        legend pos=south east,
        legend cell align={left},
        legend style={fill opacity=0.6, draw opacity=1.0, text opacity=1.0, font=\small}
        ]
        
        \addplot[black, solid, thick, mark=triangle, mark size={3.0}, mark_style
                ] 
            table [y=cdf, x=rand, col sep=comma]{\datatable};
        \addlegendentry{Random};

        \addplot[black, solid, thick, mark=pentagon, mark size={3.0}, mark_style
                ] 
            table [y=cdf, x=greedy, col sep=comma]{\datatable};
        \addlegendentry{Oracle};

        \addplot[black, solid, thick, 
                mark=diamond, mark size={3.0}, mark size={3.0}, mark_style
                ] 
            table [y=cdf, x=sweep24, col sep=comma]{\datatable};
        \addlegendentry{BS-sweep};

        \addplot[black, solid, thick, 
                mark=star, mark size={3.0}, mark size={3.0}, mark_style
                ] 
            table [y=cdf, x=ddpg_df_2x128_gamma060_000, col sep=comma]{\datatable};
        \addlegendentry{Vanilla DDPG};

        \addplot[red, solid, thick, 
                mark=diamond, mark size={3.0}, mark size={3.0}, mark_style
                ] 
            table [y=cdf, x=ddpg_2x128_gamma060_000, col sep=comma]{\datatablenancyvarsigma};
        \addlegendentry{Proposed};
        
        \addplot[green, solid, thick, 
                mark=diamond, mark size={3.0}, mark size={3.0}, mark_style
                ] 
            table [y=cdf, x=ddpg_2x128_gamma060_000, col sep=comma]{\datatablenancybsknownratecdf};
        \addlegendentry{BSOracle};
        
        \addplot[blue, solid, thick, 
                mark=diamond, mark size={3.0}, mark size={3.0}, mark_style
                ] 
            table [y=cdf, x=ddpg_2x128_gamma060_000, col sep=comma]{\datatablenancyangleknownratecdf};
        \addlegendentry{AngleOracle};
    \end{axis}
\end{tikzpicture}
        }
        \caption{$N_t = 8 \times 8$}
        \label{fig:ratecdf_03ue_bs08x08}
    \end{subfigure}%
    \begin{subfigure}{.33\linewidth}
        \resizebox{\linewidth}{!}{
            \pgfplotstableread[col sep = comma]{./data/00_Nancy/data_03UE_BS16x16_rateCdf_Baselines_NANZ.csv}\datatable
            \pgfplotstableread[col sep = comma]{./data/02_psuedo_act/data_03UE_BS16x16_rateCdfv1.csv}\tabpsuedoactvone
            \pgfplotstableread[col sep = comma]{./data/00_Nancy/data_03UE_BS16x16_rateCdfNANZ-BS-noisyaction21.csv}\datatablenancyvarsigma
            \pgfplotstableread[col sep = comma]{./data/00_Nancy/data_03UE_BS16x16_rateCdfNANZ-BSKnown.csv}\datatablenancybsknownratecdf
            \pgfplotstableread[col sep = comma]{./data/00_Nancy/data_03UE_BS16x16_rateCdfNANZ-AngleKnown.csv}\datatablenancyangleknownratecdf
            \pgfplotsset{ignore legend}
            \tikzstyle{mark_style} = [mark size={3.0}, mark repeat=20, mark phase=1]
\begin{tikzpicture}[thick,scale=0.8]
    \begin{axis}[
        width=8cm,
        height=6cm,
        xmin=-1.0,
        xmax=25,
        ymin=-0.1,
        ymax=+1.1,
        grid=major,
        xlabel={Rate (bits/sec/Hz)},
        ylabel={CDF},
        xlabel style={at={(0.50,0.05)}},
        ylabel style={at={(0.06,0.50)}},
        ytick={0.0,0.2,...,1.0},
        legend pos=south east,
        legend cell align={left},
        legend style={fill opacity=0.6, draw opacity=1.0, text opacity=1.0, font=\small}
        ]
        
        \addplot[black, solid, thick, mark=triangle, mark size={3.0}, mark_style
                ] 
            table [y=cdf, x=rand, col sep=comma]{\datatable};
        \addlegendentry{Random};

        \addplot[black, solid, thick, mark=pentagon, mark size={3.0}, mark_style
                ] 
            table [y=cdf, x=greedy, col sep=comma]{\datatable};
        \addlegendentry{Oracle};

        \addplot[black, solid, thick, 
                mark=diamond, mark size={3.0}, mark size={3.0}, mark_style
                ] 
            table [y=cdf, x=sweep24, col sep=comma]{\datatable};
        \addlegendentry{BS-sweep};

        \addplot[black, solid, thick, 
                mark=star, mark size={3.0}, mark size={3.0}, mark_style
                ] 
            table [y=cdf, x=ddpg_df_2x128_gamma060_000, col sep=comma]{\datatable};
        \addlegendentry{Vanilla DDPG};

        \addplot[red, solid, thick, 
                mark=diamond, mark size={3.0}, mark size={3.0}, mark_style
                ] 
            table [y=cdf, x=ddpg_2x128_gamma060_000, col sep=comma]{\datatablenancyvarsigma};
        \addlegendentry{Proposed};
        
        \addplot[green, solid, thick, 
                mark=diamond, mark size={3.0}, mark size={3.0}, mark_style
                ] 
            table [y=cdf, x=ddpg_2x128_gamma060_000, col sep=comma]{\datatablenancybsknownratecdf};
        \addlegendentry{BSOracle};
        
        \addplot[blue, solid, thick, 
                mark=diamond, mark size={3.0}, mark size={3.0}, mark_style
                ] 
            table [y=cdf, x=ddpg_2x128_gamma060_000, col sep=comma]{\datatablenancyangleknownratecdf};
        \addlegendentry{AngleOracle};
    \end{axis}
\end{tikzpicture}
        }
        \caption{$N_t = 16 \times 16$}
        \label{fig:ratecdf_03ue_bs16x16}
    \end{subfigure}%
    \caption{CDF of observed rates for $N_{UE} = 3$.}
    \label{fig:ratecdf_03ue}
\end{figure*}
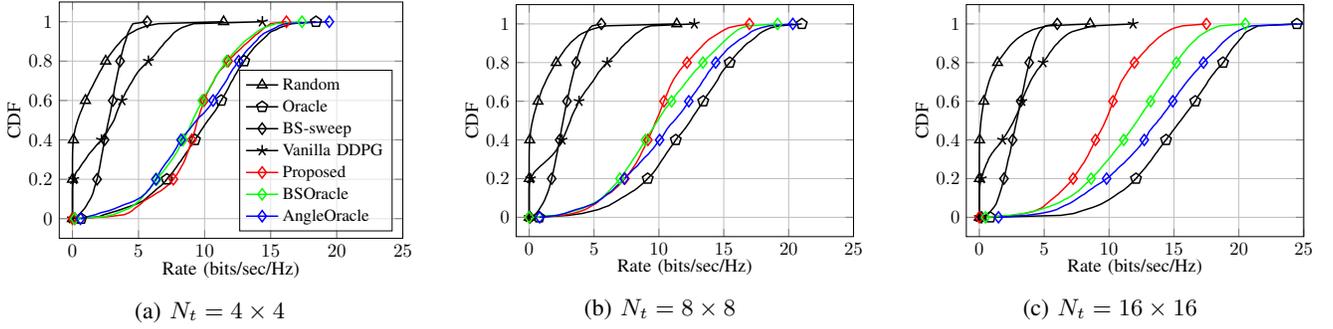

\begin{figure*}[!t]
    \centering
    \begin{subfigure}{.33\linewidth}
        \resizebox{\linewidth}{!}{
            \pgfplotstableread[col sep = comma]{./data/00_Nancy/data_05UE_BS04x04_rateCdf_Baselines_NANZ.csv}\datatable
            \pgfplotstableread[col sep = comma]{./data/02_psuedo_act/data_05UE_BS04x04_rateCdfv1.csv}\tabpsuedoactvone
            \pgfplotstableread[col sep = comma]{./data/00_Nancy/data_05UE_BS04x04_rateCdfNANZ-BS-noisyaction21.csv}\datatablenancyvarsigma
            \pgfplotstableread[col sep = comma]{./data/00_Nancy/data_05UE_BS04x04_rateCdfNANZ-BSKnown.csv}\datatablenancybsknownratecdf
            \pgfplotstableread[col sep = comma]{./data/00_Nancy/data_05UE_BS04x04_rateCdfNANZ-AngleKnown.csv}\datatablenancyangleknownratecdf
            \tikzstyle{mark_style} = [mark size={3.0}, mark repeat=20, mark phase=1]
\begin{tikzpicture}[thick,scale=0.8]
    \begin{axis}[
        width=8cm,
        height=6cm,
        xmin=-1.0,
        xmax=25,
        ymin=-0.1,
        ymax=+1.1,
        grid=major,
        xlabel={Rate (bits/sec/Hz)},
        ylabel={CDF},
        xlabel style={at={(0.50,0.05)}},
        ylabel style={at={(0.06,0.50)}},
        ytick={0.0,0.2,...,1.0},
        legend pos=south east,
        legend cell align={left},
        legend style={fill opacity=0.6, draw opacity=1.0, text opacity=1.0, font=\small}
        ]
        
        \addplot[black, solid, thick, mark=triangle, mark size={3.0}, mark_style
                ] 
            table [y=cdf, x=rand, col sep=comma]{\datatable};
        \addlegendentry{Random};

        \addplot[black, solid, thick, mark=pentagon, mark size={3.0}, mark_style
                ] 
            table [y=cdf, x=greedy, col sep=comma]{\datatable};
        \addlegendentry{Oracle};

        \addplot[black, solid, thick, 
                mark=diamond, mark size={3.0}, mark size={3.0}, mark_style
                ] 
            table [y=cdf, x=sweep24, col sep=comma]{\datatable};
        \addlegendentry{BS-sweep};

        \addplot[black, solid, thick, 
                mark=star, mark size={3.0}, mark size={3.0}, mark_style
                ] 
            table [y=cdf, x=ddpg_df_2x128_gamma060_000, col sep=comma]{\datatable};
        \addlegendentry{Vanilla DDPG};

        \addplot[red, solid, thick, 
                mark=diamond, mark size={3.0}, mark size={3.0}, mark_style
                ] 
            table [y=cdf, x=ddpg_2x128_gamma060_000, col sep=comma]{\datatablenancyvarsigma};
        \addlegendentry{Proposed};
        
        \addplot[green, solid, thick, 
                mark=diamond, mark size={3.0}, mark size={3.0}, mark_style
                ] 
            table [y=cdf, x=ddpg_2x128_gamma060_000, col sep=comma]{\datatablenancybsknownratecdf};
        \addlegendentry{BSOracle};
        
        \addplot[blue, solid, thick, 
                mark=diamond, mark size={3.0}, mark size={3.0}, mark_style
                ] 
            table [y=cdf, x=ddpg_2x128_gamma060_000, col sep=comma]{\datatablenancyangleknownratecdf};
        \addlegendentry{AngleOracle};
    \end{axis}
\end{tikzpicture}
        }
        \caption{$N_t = 4 \times 4$}
        \label{fig:ratecdf_05ue_bs04x04}
    \end{subfigure}%
    \begin{subfigure}{.33\linewidth}
        \resizebox{\linewidth}{!}{
            \pgfplotstableread[col sep = comma]{./data/00_Nancy/data_05UE_BS08x08_rateCdf_Baselines_NANZ.csv}\datatable
            \pgfplotstableread[col sep = comma]{./data/02_psuedo_act/data_05UE_BS08x08_rateCdfv1.csv}\tabpsuedoactvone
            \pgfplotstableread[col sep = comma]{./data/00_Nancy/data_05UE_BS08x08_rateCdfNANZ-BS-noisyaction21.csv}\datatablenancyvarsigma
            \pgfplotstableread[col sep = comma]{./data/00_Nancy/data_05UE_BS08x08_rateCdfNANZ-BSKnown.csv}\datatablenancybsknownratecdf
            \pgfplotstableread[col sep = comma]{./data/00_Nancy/data_05UE_BS08x08_rateCdfNANZ-AngleKnown.csv}\datatablenancyangleknownratecdf
            \pgfplotsset{ignore legend}
            \tikzstyle{mark_style} = [mark size={3.0}, mark repeat=20, mark phase=1]
\begin{tikzpicture}[thick,scale=0.8]
    \begin{axis}[
        width=8cm,
        height=6cm,
        xmin=-1.0,
        xmax=25,
        ymin=-0.1,
        ymax=+1.1,
        grid=major,
        xlabel={Rate (bits/sec/Hz)},
        ylabel={CDF},
        xlabel style={at={(0.50,0.05)}},
        ylabel style={at={(0.06,0.50)}},
        ytick={0.0,0.2,...,1.0},
        legend pos=south east,
        legend cell align={left},
        legend style={fill opacity=0.6, draw opacity=1.0, text opacity=1.0, font=\small}
        ]
        
        \addplot[black, solid, thick, mark=triangle, mark size={3.0}, mark_style
                ] 
            table [y=cdf, x=rand, col sep=comma]{\datatable};
        \addlegendentry{Random};

        \addplot[black, solid, thick, mark=pentagon, mark size={3.0}, mark_style
                ] 
            table [y=cdf, x=greedy, col sep=comma]{\datatable};
        \addlegendentry{Oracle};

        \addplot[black, solid, thick, 
                mark=diamond, mark size={3.0}, mark size={3.0}, mark_style
                ] 
            table [y=cdf, x=sweep24, col sep=comma]{\datatable};
        \addlegendentry{BS-sweep};

        \addplot[black, solid, thick, 
                mark=star, mark size={3.0}, mark size={3.0}, mark_style
                ] 
            table [y=cdf, x=ddpg_df_2x128_gamma060_000, col sep=comma]{\datatable};
        \addlegendentry{Vanilla DDPG};

        \addplot[red, solid, thick, 
                mark=diamond, mark size={3.0}, mark size={3.0}, mark_style
                ] 
            table [y=cdf, x=ddpg_2x128_gamma060_000, col sep=comma]{\datatablenancyvarsigma};
        \addlegendentry{Proposed};
        
        \addplot[green, solid, thick, 
                mark=diamond, mark size={3.0}, mark size={3.0}, mark_style
                ] 
            table [y=cdf, x=ddpg_2x128_gamma060_000, col sep=comma]{\datatablenancybsknownratecdf};
        \addlegendentry{BSOracle};
        
        \addplot[blue, solid, thick, 
                mark=diamond, mark size={3.0}, mark size={3.0}, mark_style
                ] 
            table [y=cdf, x=ddpg_2x128_gamma060_000, col sep=comma]{\datatablenancyangleknownratecdf};
        \addlegendentry{AngleOracle};
    \end{axis}
\end{tikzpicture}
        }
        
        \caption{$N_t = 8 \times 8$}
        \label{fig:ratecdf_05ue_bs08x08}
    \end{subfigure}%
    \begin{subfigure}{.33\linewidth}
        \resizebox{\linewidth}{!}{
            \pgfplotstableread[col sep = comma]{./data/00_Nancy/data_05UE_BS16x16_rateCdf_Baselines_NANZ.csv}\datatable
            \pgfplotstableread[col sep = comma]{./data/02_psuedo_act/data_05UE_BS16x16_rateCdfv1.csv}\tabpsuedoactvone
            
            \pgfplotstableread[col sep = comma]{./data/00_Nancy/data_05UE_BS16x16_rateCdfNANZ-BS-noisyaction21.csv}\datatablenancyvarsigma
            \pgfplotstableread[col sep = comma]{./data/00_Nancy/data_05UE_BS16x16_rateCdfNANZ-BSKnown.csv}\datatablenancybsknownratecdf
            \pgfplotstableread[col sep = comma]{./data/00_Nancy/data_05UE_BS16x16_rateCdfNANZ-BSKnown.csv}\datatablenancybsknownratecdf
            \pgfplotstableread[col sep = comma]{./data/00_Nancy/data_05UE_BS16x16_rateCdfNANZ-AngleKnown.csv}\datatablenancyangleknownratecdf
            \pgfplotsset{ignore legend}
            \tikzstyle{mark_style} = [mark size={3.0}, mark repeat=20, mark phase=1]
\begin{tikzpicture}[thick,scale=0.8]
    \begin{axis}[
        width=8cm,
        height=6cm,
        xmin=-1.0,
        xmax=25,
        ymin=-0.1,
        ymax=+1.1,
        grid=major,
        xlabel={Rate (bits/sec/Hz)},
        ylabel={CDF},
        xlabel style={at={(0.50,0.05)}},
        ylabel style={at={(0.06,0.50)}},
        ytick={0.0,0.2,...,1.0},
        legend pos=south east,
        legend cell align={left},
        legend style={fill opacity=0.6, draw opacity=1.0, text opacity=1.0, font=\small}
        ]
        
        \addplot[black, solid, thick, mark=triangle, mark size={3.0}, mark_style
                ] 
            table [y=cdf, x=rand, col sep=comma]{\datatable};
        \addlegendentry{Random};

        \addplot[black, solid, thick, mark=pentagon, mark size={3.0}, mark_style
                ] 
            table [y=cdf, x=greedy, col sep=comma]{\datatable};
        \addlegendentry{Oracle};

        \addplot[black, solid, thick, 
                mark=diamond, mark size={3.0}, mark size={3.0}, mark_style
                ] 
            table [y=cdf, x=sweep24, col sep=comma]{\datatable};
        \addlegendentry{BS-sweep};

        \addplot[black, solid, thick, 
                mark=star, mark size={3.0}, mark size={3.0}, mark_style
                ] 
            table [y=cdf, x=ddpg_df_2x128_gamma060_000, col sep=comma]{\datatable};
        \addlegendentry{Vanilla DDPG};

        \addplot[red, solid, thick, 
                mark=diamond, mark size={3.0}, mark size={3.0}, mark_style
                ] 
            table [y=cdf, x=ddpg_2x128_gamma060_000, col sep=comma]{\datatablenancyvarsigma};
        \addlegendentry{Proposed};
        
        \addplot[green, solid, thick, 
                mark=diamond, mark size={3.0}, mark size={3.0}, mark_style
                ] 
            table [y=cdf, x=ddpg_2x128_gamma060_000, col sep=comma]{\datatablenancybsknownratecdf};
        \addlegendentry{BSOracle};
        
        \addplot[blue, solid, thick, 
                mark=diamond, mark size={3.0}, mark size={3.0}, mark_style
                ] 
            table [y=cdf, x=ddpg_2x128_gamma060_000, col sep=comma]{\datatablenancyangleknownratecdf};
        \addlegendentry{AngleOracle};
    \end{axis}
\end{tikzpicture}
        }
        \caption{$N_t = 16 \times 16$}
        \label{fig:ratecdf_05ue_bs16x16}
    \end{subfigure}%
    \caption{CDF of observed rates for $N_{UE} = 5$.}
    \label{fig:ratecdf_05ue}
\end{figure*}
\fi

\ifCLASSOPTIONonecolumn
    \begin{figure*}[t]
        \centering
        \begin{subfigure}{.33\linewidth}
            \resizebox{\linewidth}{!}{
                \pgfplotstableread[col sep = comma]{./data/00_Nancy/data_03UE_BS04x04_rateEvo_Baselines_NANZ.csv}\datatable
                \pgfplotstableread[col sep = comma]{./data/00_Nancy/data_03UE_BS04x04_rateEvoNANZ-BS-noisyaction21.csv}\datatablenancysigmaone 
                \pgfplotstableread[col sep = comma]{./data/00_Nancy/data_03UE_BS04x04_rateEvoNANZ-BSKnown.csv}\datatablenancybsknownrateevo
               \pgfplotstableread[col sep = comma]{./data/00_Nancy/data_03UE_BS04x04_rateEvoNANZ-AngleKnown.csv}\datatablenancyangleknownrateevo
            \tikzstyle{mark_style} = []

\begin{tikzpicture}[thick,scale=0.8]
    \begin{semilogyaxis}[
        width=8cm,
        height=6cm,
        xmin=0,
        xmax=7000,
        ymin=0.5,
        ymax=20,
        grid=major,
        xlabel={No. of Episodes},
        ylabel={Avg. Sum Rate (bits/sec/Hz)},
        xlabel style={at={(0.50,0.05)}},
        ylabel style={at={(0.06,0.50)}},
        ytick={0,1,10},
        xtick={0,2000, 4000, 6000, 7000},
        legend pos=south east,
        legend cell align={left},
        legend style={fill opacity=0.6, draw opacity=1.0, text opacity=1.0, font=\small}
        ]
        
        \addplot[black, solid, thick, mark=triangle, mark size={3.0}, mark_style, mark repeat=3,
                ] 
            table [x=x_data, y expr=\thisrow{y_rand}/1000, col sep=comma]{\datatable};
        \addlegendentry{Random};

        \addplot[black, solid, thick, mark=pentagon, mark size={3.0}, mark_style, mark repeat=3,
                ] 
            table [x=x_data, y expr=\thisrow{y_greedy}/1000, col sep=comma]{\datatable};
        \addlegendentry{Oracle};

        \addplot[black, solid, thick, 
                mark=diamond, mark size={3.0}, mark size={3.0}, mark_style, mark repeat=3,
                ] 
            table [x=x_data, y expr=\thisrow{y_sweep24}/1000, col sep=comma]{\datatable};
        \addlegendentry{BS-sweep};

        \addplot[black, solid, thick, 
                mark=star, mark size={3.0}, mark size={3.0}, mark_style, mark repeat=3,
                ] 
            table [x=x_data, y expr=\thisrow{y_ddpg_df_2x128_gamma060}/1000, col sep=comma]{\datatable};
        \addlegendentry{Vanilla DDPG};

        \addplot[red, solid, thick, 
                mark=diamond, mark size={3.0}, mark size={3.0}, mark_style, mark repeat=3,
                ] 
            table [x=x_data, y expr=\thisrow{y_ddpg_2x128_gamma060}/1000, col sep=comma]{\datatablenancysigmaone};
        \addlegendentry{Proposed};
        
        \addplot[green, solid, thick, 
                mark=diamond, mark size={3.0}, mark size={3.0}, mark_style, mark repeat=3,
                ] 
            table [x=x_data, y expr=\thisrow{y_ddpg_2x128_gamma060}/1000, col sep=comma]{\datatablenancybsknownrateevo};
        \addlegendentry{BSOracle};
        
        \addplot[blue, solid, thick, 
                mark=diamond, mark size={3.0}, mark size={3.0}, mark_style, mark repeat=3,
                ] 
            table [x=x_data, y expr=\thisrow{y_ddpg_2x128_gamma060}/1000, col sep=comma]{\datatablenancyangleknownrateevo};
        \addlegendentry{AngleOracle};
    \end{semilogyaxis}
\end{tikzpicture}
            }
            \caption{$N_t = 4 \times 4$}
            \label{fig:rateevo_03ue_bs04x04}
        \end{subfigure}%
        \begin{subfigure}{.33\linewidth}
            \resizebox{\linewidth}{!}{
                \pgfplotstableread[col sep = comma]{./data/00_Nancy/data_03UE_BS08x08_rateEvo_Baselines_NANZ.csv}\datatable
                \pgfplotstableread[col sep = comma]{./data/00_Nancy/data_03UE_BS08x08_rateEvoNANZ-BS-noisyaction21.csv}\datatablenancysigmaone
                \pgfplotstableread[col sep = comma]{./data/00_Nancy/data_03UE_BS08x08_rateEvoNANZ-BSKnown.csv}\datatablenancybsknownrateevo
                \pgfplotstableread[col sep = comma]{./data/00_Nancy/data_03UE_BS08x08_rateEvoNANZ-AngleKnown.csv}\datatablenancyangleknownrateevo
                \pgfplotsset{ignore legend}
                \tikzstyle{mark_style} = []
\begin{tikzpicture}[thick,scale=0.8]
    \begin{semilogyaxis}[
        width=8cm,
        height=6cm,
        xmin=0,
        xmax=7000,
        ymin=0.5,
        ymax=20,
        grid=major,
        xlabel={No. of Episodes},
        ylabel={Avg. Sum Rate (bits/sec/Hz)},
        xlabel style={at={(0.50,0.05)}},
        ylabel style={at={(0.06,0.50)}},
        ytick={0,1,10},
        xtick={0,2000, 4000, 6000, 7000},
        legend pos=south east,
        legend cell align={left},
        legend style={fill opacity=0.6, draw opacity=1.0, text opacity=1.0, font=\small}
        ]
        
        \addplot[black, solid, thick, mark=triangle, mark size={3.0}, mark_style, mark repeat=3,
                ] 
            table [x=x_data, y expr=\thisrow{y_rand}/1000, col sep=comma]{\datatable};
        \addlegendentry{Random};

        \addplot[black, solid, thick, mark=pentagon, mark size={3.0}, mark_style, mark repeat=3,
                ] 
            table [x=x_data, y expr=\thisrow{y_greedy}/1000, col sep=comma]{\datatable};
        \addlegendentry{Oracle};

        \addplot[black, solid, thick, 
                mark=diamond, mark size={3.0}, mark size={3.0}, mark_style, mark repeat=3,
                ] 
            table [x=x_data, y expr=\thisrow{y_sweep24}/1000, col sep=comma]{\datatable};
        \addlegendentry{BS-sweep};

        \addplot[black, solid, thick, 
                mark=star, mark size={3.0}, mark size={3.0}, mark_style, mark repeat=3,
                ] 
            table [x=x_data, y expr=\thisrow{y_ddpg_df_2x128_gamma060}/1000, col sep=comma]{\datatable};
        \addlegendentry{Vanilla DDPG};

        \addplot[red, solid, thick, 
                mark=diamond, mark size={3.0}, mark size={3.0}, mark_style, mark repeat=3,
                ] 
            table [x=x_data, y expr=\thisrow{y_ddpg_2x128_gamma060}/1000, col sep=comma]{\datatablenancysigmaone};
        \addlegendentry{Proposed};
        
        \addplot[green, solid, thick, 
                mark=diamond, mark size={3.0}, mark size={3.0}, mark_style, mark repeat=3,
                ] 
            table [x=x_data, y expr=\thisrow{y_ddpg_2x128_gamma060}/1000, col sep=comma]{\datatablenancybsknownrateevo};
        \addlegendentry{BSOracle};
        
        \addplot[blue, solid, thick, 
                mark=diamond, mark size={3.0}, mark size={3.0}, mark_style, mark repeat=3,
                ] 
            table [x=x_data, y expr=\thisrow{y_ddpg_2x128_gamma060}/1000, col sep=comma]{\datatablenancyangleknownrateevo};
        \addlegendentry{AngleOracle};
    \end{semilogyaxis}
\end{tikzpicture}
            }
            \caption{$N_t = 8 \times 8$}
            \label{fig:rateevo_03ue_bs08x08}
        \end{subfigure}%
        \begin{subfigure}{.33\linewidth}
            \resizebox{\linewidth}{!}{
                \pgfplotstableread[col sep = comma]{./data/00_Nancy/data_03UE_BS16x16_rateEvo_Baselines_NANZ.csv}\datatable
                \pgfplotstableread[col sep = comma]{./data/02_psuedo_act/data_03UE_BS16x16_rateEvov1.csv}\tabpsuedoactvone
                \pgfplotstableread[col sep = comma]{./data/00_Nancy/data_03UE_BS16x16_rateEvoNANZ-BS-noisyaction21.csv}\datatablenancysigmaone
                \pgfplotstableread[col sep = comma]{./data/00_Nancy/data_03UE_BS16x16_rateEvoNANZ-BSKnown.csv}\datatablenancybsknownrateevo
                \pgfplotstableread[col sep = comma]{./data/00_Nancy/data_03UE_BS16x16_rateEvoNANZ-AngleKnown.csv}\datatablenancyangleknownrateevo
                \pgfplotsset{ignore legend}
                \tikzstyle{mark_style} = []
\begin{tikzpicture}[thick,scale=0.8]
    \begin{semilogyaxis}[
        width=8cm,
        height=6cm,
        xmin=0,
        xmax=7000,
        ymin=0.5,
        ymax=20,
        grid=major,
        xlabel={No. of Episodes},
        ylabel={Avg. Sum Rate (bits/sec/Hz)},
        xlabel style={at={(0.50,0.05)}},
        ylabel style={at={(0.06,0.50)}},
        ytick={0,1,10},
        xtick={0,2000, 4000, 6000, 7000},
        legend pos=south east,
        legend cell align={left},
        legend style={fill opacity=0.6, draw opacity=1.0, text opacity=1.0, font=\small}
        ]
        
        \addplot[black, solid, thick, mark=triangle, mark size={3.0}, mark_style, mark repeat=3,
                ] 
            table [x=x_data, y expr=\thisrow{y_rand}/1000, col sep=comma]{\datatable};
        \addlegendentry{Random};

        \addplot[black, solid, thick, mark=pentagon, mark size={3.0}, mark_style, mark repeat=3,
                ] 
            table [x=x_data, y expr=\thisrow{y_greedy}/1000, col sep=comma]{\datatable};
        \addlegendentry{Oracle};

        \addplot[black, solid, thick, 
                mark=diamond, mark size={3.0}, mark size={3.0}, mark_style, mark repeat=3,
                ] 
            table [x=x_data, y expr=\thisrow{y_sweep24}/1000, col sep=comma]{\datatable};
        \addlegendentry{BS-sweep};

        \addplot[black, solid, thick, 
                mark=star, mark size={3.0}, mark size={3.0}, mark_style, mark repeat=3,
                ] 
            table [x=x_data, y expr=\thisrow{y_ddpg_df_2x128_gamma060}/1000, col sep=comma]{\datatable};
        \addlegendentry{Vanilla DDPG};

        \addplot[red, solid, thick, 
                mark=diamond, mark size={3.0}, mark size={3.0}, mark_style, mark repeat=3,
                ] 
            table [x=x_data, y expr=\thisrow{y_ddpg_2x128_gamma060}/1000, col sep=comma]{\datatablenancysigmaone};
        \addlegendentry{Proposed};
        
        \addplot[green, solid, thick, 
                mark=diamond, mark size={3.0}, mark size={3.0}, mark_style, mark repeat=3,
                ] 
            table [x=x_data, y expr=\thisrow{y_ddpg_2x128_gamma060}/1000, col sep=comma]{\datatablenancybsknownrateevo};
        \addlegendentry{BSOracle};
        
        \addplot[blue, solid, thick, 
                mark=diamond, mark size={3.0}, mark size={3.0}, mark_style, mark repeat=3,
                ] 
            table [x=x_data, y expr=\thisrow{y_ddpg_2x128_gamma060}/1000, col sep=comma]{\datatablenancyangleknownrateevo};
        \addlegendentry{AngleOracle};
    \end{semilogyaxis}
\end{tikzpicture}
            }
            \caption{$N_t = 16 \times 16$}
            \label{fig:rateevo_03ue_bs16x16}
        \end{subfigure}%
        \caption{Rate Evolution during learning for $N_{UE} = 3$.}
        \label{fig:rateevo_03ue}
    \end{figure*}
    
    \begin{figure*}[t]
        \centering
        \begin{subfigure}{.33\linewidth}
            \resizebox{\linewidth}{!}{
                \pgfplotstableread[col sep = comma]{./data/00_Nancy/data_05UE_BS04x04_rateEvo_Baselines_NANZ.csv}\datatable
              \pgfplotstableread[col sep = comma]{./data/00_Nancy/data_05UE_BS04x04_rateEvoNANZ-BS-noisyaction21.csv}\datatablenancysigmaone 
              \pgfplotstableread[col sep = comma]{./data/00_Nancy/data_05UE_BS04x04_rateEvoNANZ-BSKnown.csv}\datatablenancybsknownrateevo
              \pgfplotstableread[col sep = comma]{./data/00_Nancy/data_05UE_BS04x04_rateEvoNANZ-AngleKnown.csv}\datatablenancyangleknownrateevo
              \pgfplotsset{ignore legend}
            \tikzstyle{mark_style} = []
\begin{tikzpicture}[thick,scale=0.8]
    \begin{semilogyaxis}[
        width=8cm,
        height=6cm,
        xmin=0,
        xmax=7000,
        ymin=0.5,
        ymax=20,
        grid=major,
        xlabel={No. of Episodes},
        ylabel={Avg. Sum Rate (bits/sec/Hz)},
        xlabel style={at={(0.50,0.05)}},
        ylabel style={at={(0.06,0.50)}},
        ytick={0,1,10},
        xtick={0,2000, 4000, 6000, 7000},
        legend pos=south east,
        legend cell align={left},
        legend style={fill opacity=0.6, draw opacity=1.0, text opacity=1.0, font=\small}
        ]
        
        \addplot[black, solid, thick, mark=triangle, mark size={3.0}, mark_style, mark repeat=3,
                ] 
            table [x=x_data, y expr=\thisrow{y_rand}/1000, col sep=comma]{\datatable};
        \addlegendentry{Random};

        \addplot[black, solid, thick, mark=pentagon, mark size={3.0}, mark_style, mark repeat=3,
                ] 
            table [x=x_data, y expr=\thisrow{y_greedy}/1000, col sep=comma]{\datatable};
        \addlegendentry{Oracle};

        \addplot[black, solid, thick, 
                mark=diamond, mark size={3.0}, mark size={3.0}, mark_style, mark repeat=3,
                ] 
            table [x=x_data, y expr=\thisrow{y_sweep24}/1000, col sep=comma]{\datatable};
        \addlegendentry{BS-sweep};

        \addplot[black, solid, thick, 
                mark=star, mark size={3.0}, mark size={3.0}, mark_style, mark repeat=3,
                ] 
            table [x=x_data, y expr=\thisrow{y_ddpg_df_2x128_gamma060}/1000, col sep=comma]{\datatable};
        \addlegendentry{Vanilla DDPG};

        \addplot[red, solid, thick, 
                mark=diamond, mark size={3.0}, mark size={3.0}, mark_style, mark repeat=3,
                ] 
            table [x=x_data, y expr=\thisrow{y_ddpg_2x128_gamma060}/1000, col sep=comma]{\datatablenancysigmaone};
        \addlegendentry{Proposed};
        
        \addplot[green, solid, thick, 
                mark=diamond, mark size={3.0}, mark size={3.0}, mark_style, mark repeat=3,
                ] 
            table [x=x_data, y expr=\thisrow{y_ddpg_2x128_gamma060}/1000, col sep=comma]{\datatablenancybsknownrateevo};
        \addlegendentry{BSOracle};
        
        \addplot[blue, solid, thick, 
                mark=diamond, mark size={3.0}, mark size={3.0}, mark_style, mark repeat=3,
                ] 
            table [x=x_data, y expr=\thisrow{y_ddpg_2x128_gamma060}/1000, col sep=comma]{\datatablenancyangleknownrateevo};
        \addlegendentry{AngleOracle};
    \end{semilogyaxis}
\end{tikzpicture}
            }
            \caption{$N_t = 4 \times 4$}
            \label{fig:rateevo_05ue_bs04x04}
        \end{subfigure}%
        \begin{subfigure}{.33\linewidth}
            \resizebox{\linewidth}{!}{
                \pgfplotstableread[col sep = comma]{./data/00_Nancy/data_05UE_BS08x08_rateEvo_Baselines_NANZ.csv}\datatable
                \pgfplotstableread[col sep = comma]{./data/02_psuedo_act/data_05UE_BS08x08_rateEvov1.csv}\tabpsuedoactvone
                \pgfplotstableread[col sep = comma]{./data/00_Nancy/data_05UE_BS08x08_rateEvoNANZ-BS-noisyaction21.csv}\datatablenancysigmaone
                \pgfplotstableread[col sep = comma]{./data/00_Nancy/data_05UE_BS08x08_rateEvoNANZ-BSKnown.csv}\datatablenancybsknownrateevo
                \pgfplotstableread[col sep = comma]{./data/00_Nancy/data_05UE_BS08x08_rateEvoNANZ-AngleKnown.csv}\datatablenancyangleknownrateevo
                \pgfplotsset{ignore legend}
                \tikzstyle{mark_style} = []
\begin{tikzpicture}[thick,scale=0.8]
    \begin{semilogyaxis}[
        width=8cm,
        height=6cm,
        xmin=0,
        xmax=7000,
        ymin=0.5,
        ymax=20,
        grid=major,
        xlabel={No. of Episodes},
        ylabel={Avg. Sum Rate (bits/sec/Hz)},
        xlabel style={at={(0.50,0.05)}},
        ylabel style={at={(0.06,0.50)}},
        ytick={0,1,10},
        xtick={0,2000, 4000, 6000, 7000},
        legend pos=south east,
        legend cell align={left},
        legend style={fill opacity=0.6, draw opacity=1.0, text opacity=1.0, font=\small}
        ]
        
        \addplot[black, solid, thick, mark=triangle, mark size={3.0}, mark_style, mark repeat=3,
                ] 
            table [x=x_data, y expr=\thisrow{y_rand}/1000, col sep=comma]{\datatable};
        \addlegendentry{Random};

        \addplot[black, solid, thick, mark=pentagon, mark size={3.0}, mark_style, mark repeat=3,
                ] 
            table [x=x_data, y expr=\thisrow{y_greedy}/1000, col sep=comma]{\datatable};
        \addlegendentry{Oracle};

        \addplot[black, solid, thick, 
                mark=diamond, mark size={3.0}, mark size={3.0}, mark_style, mark repeat=3,
                ] 
            table [x=x_data, y expr=\thisrow{y_sweep24}/1000, col sep=comma]{\datatable};
        \addlegendentry{BS-sweep};

        \addplot[black, solid, thick, 
                mark=star, mark size={3.0}, mark size={3.0}, mark_style, mark repeat=3,
                ] 
            table [x=x_data, y expr=\thisrow{y_ddpg_df_2x128_gamma060}/1000, col sep=comma]{\datatable};
        \addlegendentry{Vanilla DDPG};

        \addplot[red, solid, thick, 
                mark=diamond, mark size={3.0}, mark size={3.0}, mark_style, mark repeat=3,
                ] 
            table [x=x_data, y expr=\thisrow{y_ddpg_2x128_gamma060}/1000, col sep=comma]{\datatablenancysigmaone};
        \addlegendentry{Proposed};
        
        \addplot[green, solid, thick, 
                mark=diamond, mark size={3.0}, mark size={3.0}, mark_style, mark repeat=3,
                ] 
            table [x=x_data, y expr=\thisrow{y_ddpg_2x128_gamma060}/1000, col sep=comma]{\datatablenancybsknownrateevo};
        \addlegendentry{BSOracle};
        
        \addplot[blue, solid, thick, 
                mark=diamond, mark size={3.0}, mark size={3.0}, mark_style, mark repeat=3,
                ] 
            table [x=x_data, y expr=\thisrow{y_ddpg_2x128_gamma060}/1000, col sep=comma]{\datatablenancyangleknownrateevo};
        \addlegendentry{AngleOracle};
    \end{semilogyaxis}
\end{tikzpicture}
            }
            \caption{$N_t = 8 \times 8$}
            \label{fig:rateevo_05ue_bs08x08}
        \end{subfigure}%
        \begin{subfigure}{.33\linewidth}
            \resizebox{\linewidth}{!}{
                \pgfplotstableread[col sep = comma]{./data/00_Nancy/data_05UE_BS16x16_rateEvo_Baselines_NANZ.csv}\datatable
                
                \pgfplotstableread[col sep = comma]{./data/00_Nancy/data_05UE_BS16x16_rateEvoNANZ-BS-noisyaction21.csv}\datatablenancysigmaone
                
                \pgfplotstableread[col sep = comma]{./data/00_Nancy/data_05UE_BS16x16_rateEvoNANZ-BSKnown.csv}\datatablenancybsknownrateevo
                \pgfplotstableread[col sep = comma]{./data/00_Nancy/data_05UE_BS16x16_rateEvoNANZ-AngleKnown.csv}\datatablenancyangleknownrateevo
                \pgfplotsset{ignore legend}
                \tikzstyle{mark_style} = []
\begin{tikzpicture}[thick,scale=0.8]
    \begin{semilogyaxis}[
        width=8cm,
        height=6cm,
        xmin=0,
        xmax=7000,
        ymin=0.5,
        ymax=20,
        grid=major,
        xlabel={No. of Episodes},
        ylabel={Avg. Sum Rate (bits/sec/Hz)},
        xlabel style={at={(0.50,0.05)}},
        ylabel style={at={(0.06,0.50)}},
        ytick={0,1,10},
        xtick={0,2000, 4000, 6000, 7000},
        legend pos=south east,
        legend cell align={left},
        legend style={fill opacity=0.6, draw opacity=1.0, text opacity=1.0, font=\small}
        ]
        
        \addplot[black, solid, thick, mark=triangle, mark size={3.0}, mark_style, mark repeat=3,
                ] 
            table [x=x_data, y expr=\thisrow{y_rand}/1000, col sep=comma]{\datatable};
        \addlegendentry{Random};

        \addplot[black, solid, thick, mark=pentagon, mark size={3.0}, mark_style, mark repeat=3,
                ] 
            table [x=x_data, y expr=\thisrow{y_greedy}/1000, col sep=comma]{\datatable};
        \addlegendentry{Oracle};

        \addplot[black, solid, thick, 
                mark=diamond, mark size={3.0}, mark size={3.0}, mark_style, mark repeat=3,
                ] 
            table [x=x_data, y expr=\thisrow{y_sweep24}/1000, col sep=comma]{\datatable};
        \addlegendentry{BS-sweep};

        \addplot[black, solid, thick, 
                mark=star, mark size={3.0}, mark size={3.0}, mark_style, mark repeat=3,
                ] 
            table [x=x_data, y expr=\thisrow{y_ddpg_df_2x128_gamma060}/1000, col sep=comma]{\datatable};
        \addlegendentry{Vanilla DDPG};

        \addplot[red, solid, thick, 
                mark=diamond, mark size={3.0}, mark size={3.0}, mark_style, mark repeat=3,
                ] 
            table [x=x_data, y expr=\thisrow{y_ddpg_2x128_gamma060}/1000, col sep=comma]{\datatablenancysigmaone};
        \addlegendentry{Proposed};
        
        \addplot[green, solid, thick, 
                mark=diamond, mark size={3.0}, mark size={3.0}, mark_style, mark repeat=3,
                ] 
            table [x=x_data, y expr=\thisrow{y_ddpg_2x128_gamma060}/1000, col sep=comma]{\datatablenancybsknownrateevo};
        \addlegendentry{BSOracle};
        
        \addplot[blue, solid, thick, 
                mark=diamond, mark size={3.0}, mark size={3.0}, mark_style, mark repeat=3,
                ] 
            table [x=x_data, y expr=\thisrow{y_ddpg_2x128_gamma060}/1000, col sep=comma]{\datatablenancyangleknownrateevo};
        \addlegendentry{AngleOracle};
    \end{semilogyaxis}
\end{tikzpicture}
            }
            \caption{$N_t = 16 \times 16$}
            \label{fig:rateevo_05ue_bs16x16}
        \end{subfigure}%
        \caption{Rate Evolution during learning for $N_{UE} = 5$.}
        \label{fig:rateevo_05ue}
    \end{figure*}
\fi

The average sum rate evolution during the learning phase of proposed algorithm is given in Fig. \ref{fig:rateevo_03ue} (for $3$ UEs) and Fig. \ref{fig:rateevo_05ue} (for $5$ UEs) for different number of transmit antenna elements. With a more number of antenna elements, the beam produced by the UPA antenna becomes narrower thus delivering most of the power towards the target direction. With a fewer number of transmit antennas, the beams become broader and this can cause additional interference to neighboring UEs even with good spatial separation. This is evident from the trend of the average sum rate across the different number of transmitting elements. As the number of transmitting elements in the antenna increases, the SINR improves and we can see an improvement in the rate of the Oracle.  As the beams become narrower, the BS-Sweep method improves the rate.

The learning phase of all the DRL algorithms i.e. Vanilla DDPG, BSOracle, AngleOracle, and the Proposed method is evident from Fig. \ref{fig:rateevo_03ue} and Fig. \ref{fig:rateevo_05ue} (legends are shared across all the figures). The performances of all these algorithms are upper bounded by the performance of the Oracle method where the agent has all the information regarding the best $\mu BS$ and UE locations. Realizing an Oracle or even BSOracle/AngleOracle agents in a practical scenario is not possible. However, the aim is to reach as close to these as possible. Initially, both the Vanilla DDPG and the proposed learning agents start with a performance similar to that of \emph{Random} agent. This is expected as at the beginning, DRL agents are initialized with random weights and hence actions taken by them are also random. The BSOracle agent knows the best $\mu BS$ for each UE and needs to predict just the UE locations. Therefore the performance of BSOracle starts with a comparatively higher value of sum rate than Vanilla DDPG or the proposed one. However, slowly the proposed method catches up with the performance of BSOracle. The AngleOracle agent predicts the best $\mu BS$ for each UE and uses the corresponding angle predictions by an oracle. As the average sum rate is more sensitive towards UE locations than the knowledge of best $\mu BS$, the performance of the agent in the AngleOracle method starts from a better average sum rate value than the BSOracle method and it also reaches a higher average sum rate value than all other algorithms (except Oracle). Reiterating here the fact that, the proposed algorithm does not take any help from the Oracle and thus learns both the best $\mu BS$ and the angles by itself. Compared to this, the BSOracle/AngleOracle methods know a part of the information(either $\mu BS$ or UE locations) and thus can be thought of as \emph{partially-Oracle}. As training progresses, we can observe that the performance of all the DRL methods increase. The rate evolution clearly shows the advantage of the proposed neural architecture and the action space exploration for both $\mu BS$ and the elevation/azimuth angles. Since Vanilla DDPG is not able to handle the mixed actions well, the learning curve flattens soon after the start of the training and also at levels at or below that of the existing beam sweeping methods. However, the proposed architecture for action selection can overcome this difficulty and the learning progresses at a much faster pace than Vanilla DDPG. Further, the evolution of rate is slower because of the exploration in both the action spaces. After approximately $500$ episodes of training itself, the proposed method can perform better than beam sweeping methods. In all of the antenna configurations and UE configurations, at the end of the rate evolution process, the proposed method performs similar to the BSOracle method. This implies that the proposed method can find the best $\mu BS$ for each UE efficiently. With less number of UEs (3 UE case), the performance of the proposed method is very close to the Oracle. For a particular antenna dimension, the gap between the BSOracle/AngleOracle method and the Oracle increases with more number of UEs because the dimension of the action for the agent itself increases, hence increasing the problem complexity.

\ifCLASSOPTIONonecolumn
\begin{figure*}[t]
    \centering
    \begin{subfigure}{.33\linewidth}
        \resizebox{\linewidth}{!}{
            \pgfplotstableread[col sep = comma]{./data/00_Nancy/data_03UE_BS04x04_rateCdf_Baselines_NANZ.csv}\datatable
         
            \pgfplotstableread[col sep = comma]{./data/00_Nancy/data_03UE_BS04x04_rateCdfNANZ-BS-noisyaction21.csv}\datatablenancyvarsigma
            \pgfplotstableread[col sep = comma]{./data/00_Nancy/data_03UE_BS04x04_rateCdfNANZ-BSKnown.csv}\datatablenancybsknownratecdf
            \pgfplotstableread[col sep = comma]{./data/00_Nancy/data_03UE_BS04x04_rateCdfNANZ-AngleKnown.csv}\datatablenancyangleknownratecdf
            \tikzstyle{mark_style} = [mark size={3.0}, mark repeat=20, mark phase=1]
\begin{tikzpicture}[thick,scale=0.8]
    \begin{axis}[
        width=8cm,
        height=6cm,
        xmin=-1.0,
        xmax=25,
        ymin=-0.1,
        ymax=+1.1,
        grid=major,
        xlabel={Rate (bits/sec/Hz)},
        ylabel={CDF},
        xlabel style={at={(0.50,0.05)}},
        ylabel style={at={(0.06,0.50)}},
        ytick={0.0,0.2,...,1.0},
        legend pos=south east,
        legend cell align={left},
        legend style={fill opacity=0.6, draw opacity=1.0, text opacity=1.0, font=\small}
        ]
        
        \addplot[black, solid, thick, mark=triangle, mark size={3.0}, mark_style
                ] 
            table [y=cdf, x=rand, col sep=comma]{\datatable};
        \addlegendentry{Random};

        \addplot[black, solid, thick, mark=pentagon, mark size={3.0}, mark_style
                ] 
            table [y=cdf, x=greedy, col sep=comma]{\datatable};
        \addlegendentry{Oracle};

        \addplot[black, solid, thick, 
                mark=diamond, mark size={3.0}, mark size={3.0}, mark_style
                ] 
            table [y=cdf, x=sweep24, col sep=comma]{\datatable};
        \addlegendentry{BS-sweep};

        \addplot[black, solid, thick, 
                mark=star, mark size={3.0}, mark size={3.0}, mark_style
                ] 
            table [y=cdf, x=ddpg_df_2x128_gamma060_000, col sep=comma]{\datatable};
        \addlegendentry{Vanilla DDPG};

        \addplot[red, solid, thick, 
                mark=diamond, mark size={3.0}, mark size={3.0}, mark_style
                ] 
            table [y=cdf, x=ddpg_2x128_gamma060_000, col sep=comma]{\datatablenancyvarsigma};
        \addlegendentry{Proposed};
        
        \addplot[green, solid, thick, 
                mark=diamond, mark size={3.0}, mark size={3.0}, mark_style
                ] 
            table [y=cdf, x=ddpg_2x128_gamma060_000, col sep=comma]{\datatablenancybsknownratecdf};
        \addlegendentry{BSOracle};
        
        \addplot[blue, solid, thick, 
                mark=diamond, mark size={3.0}, mark size={3.0}, mark_style
                ] 
            table [y=cdf, x=ddpg_2x128_gamma060_000, col sep=comma]{\datatablenancyangleknownratecdf};
        \addlegendentry{AngleOracle};
    \end{axis}
\end{tikzpicture}
        }
        \caption{$N_t = 4 \times 4$}
        \label{fig:ratecdf_03ue_bs04x04}
    \end{subfigure}%
    \begin{subfigure}{.33\linewidth}
        \resizebox{\linewidth}{!}{
            \pgfplotstableread[col sep = comma]{./data/00_Nancy/data_03UE_BS08x08_rateCdf_Baselines_NANZ.csv}\datatable
            \pgfplotstableread[col sep = comma]{./data/00_Nancy/data_03UE_BS08x08_rateCdfNANZ-BS-noisyaction21.csv}\datatablenancyvarsigma
            \pgfplotstableread[col sep = comma]{./data/00_Nancy/data_03UE_BS08x08_rateCdfNANZ-BSKnown.csv}\datatablenancybsknownratecdf
            \pgfplotstableread[col sep = comma]{./data/00_Nancy/data_03UE_BS08x08_rateCdfNANZ-AngleKnown.csv}\datatablenancyangleknownratecdf
            \pgfplotsset{ignore legend}
            \tikzstyle{mark_style} = [mark size={3.0}, mark repeat=20, mark phase=1]
\begin{tikzpicture}[thick,scale=0.8]
    \begin{axis}[
        width=8cm,
        height=6cm,
        xmin=-1.0,
        xmax=25,
        ymin=-0.1,
        ymax=+1.1,
        grid=major,
        xlabel={Rate (bits/sec/Hz)},
        ylabel={CDF},
        xlabel style={at={(0.50,0.05)}},
        ylabel style={at={(0.06,0.50)}},
        ytick={0.0,0.2,...,1.0},
        legend pos=south east,
        legend cell align={left},
        legend style={fill opacity=0.6, draw opacity=1.0, text opacity=1.0, font=\small}
        ]
        
        \addplot[black, solid, thick, mark=triangle, mark size={3.0}, mark_style
                ] 
            table [y=cdf, x=rand, col sep=comma]{\datatable};
        \addlegendentry{Random};

        \addplot[black, solid, thick, mark=pentagon, mark size={3.0}, mark_style
                ] 
            table [y=cdf, x=greedy, col sep=comma]{\datatable};
        \addlegendentry{Oracle};

        \addplot[black, solid, thick, 
                mark=diamond, mark size={3.0}, mark size={3.0}, mark_style
                ] 
            table [y=cdf, x=sweep24, col sep=comma]{\datatable};
        \addlegendentry{BS-sweep};

        \addplot[black, solid, thick, 
                mark=star, mark size={3.0}, mark size={3.0}, mark_style
                ] 
            table [y=cdf, x=ddpg_df_2x128_gamma060_000, col sep=comma]{\datatable};
        \addlegendentry{Vanilla DDPG};

        \addplot[red, solid, thick, 
                mark=diamond, mark size={3.0}, mark size={3.0}, mark_style
                ] 
            table [y=cdf, x=ddpg_2x128_gamma060_000, col sep=comma]{\datatablenancyvarsigma};
        \addlegendentry{Proposed};
        
        \addplot[green, solid, thick, 
                mark=diamond, mark size={3.0}, mark size={3.0}, mark_style
                ] 
            table [y=cdf, x=ddpg_2x128_gamma060_000, col sep=comma]{\datatablenancybsknownratecdf};
        \addlegendentry{BSOracle};
        
        \addplot[blue, solid, thick, 
                mark=diamond, mark size={3.0}, mark size={3.0}, mark_style
                ] 
            table [y=cdf, x=ddpg_2x128_gamma060_000, col sep=comma]{\datatablenancyangleknownratecdf};
        \addlegendentry{AngleOracle};
    \end{axis}
\end{tikzpicture}
        }
        \caption{$N_t = 8 \times 8$}
        \label{fig:ratecdf_03ue_bs08x08}
    \end{subfigure}%
    \begin{subfigure}{.33\linewidth}
        \resizebox{\linewidth}{!}{
            \pgfplotstableread[col sep = comma]{./data/00_Nancy/data_03UE_BS16x16_rateCdf_Baselines_NANZ.csv}\datatable
            \pgfplotstableread[col sep = comma]{./data/02_psuedo_act/data_03UE_BS16x16_rateCdfv1.csv}\tabpsuedoactvone
            \pgfplotstableread[col sep = comma]{./data/00_Nancy/data_03UE_BS16x16_rateCdfNANZ-BS-noisyaction21.csv}\datatablenancyvarsigma
            \pgfplotstableread[col sep = comma]{./data/00_Nancy/data_03UE_BS16x16_rateCdfNANZ-BSKnown.csv}\datatablenancybsknownratecdf
            \pgfplotstableread[col sep = comma]{./data/00_Nancy/data_03UE_BS16x16_rateCdfNANZ-AngleKnown.csv}\datatablenancyangleknownratecdf
            \pgfplotsset{ignore legend}
            \tikzstyle{mark_style} = [mark size={3.0}, mark repeat=20, mark phase=1]
\begin{tikzpicture}[thick,scale=0.8]
    \begin{axis}[
        width=8cm,
        height=6cm,
        xmin=-1.0,
        xmax=25,
        ymin=-0.1,
        ymax=+1.1,
        grid=major,
        xlabel={Rate (bits/sec/Hz)},
        ylabel={CDF},
        xlabel style={at={(0.50,0.05)}},
        ylabel style={at={(0.06,0.50)}},
        ytick={0.0,0.2,...,1.0},
        legend pos=south east,
        legend cell align={left},
        legend style={fill opacity=0.6, draw opacity=1.0, text opacity=1.0, font=\small}
        ]
        
        \addplot[black, solid, thick, mark=triangle, mark size={3.0}, mark_style
                ] 
            table [y=cdf, x=rand, col sep=comma]{\datatable};
        \addlegendentry{Random};

        \addplot[black, solid, thick, mark=pentagon, mark size={3.0}, mark_style
                ] 
            table [y=cdf, x=greedy, col sep=comma]{\datatable};
        \addlegendentry{Oracle};

        \addplot[black, solid, thick, 
                mark=diamond, mark size={3.0}, mark size={3.0}, mark_style
                ] 
            table [y=cdf, x=sweep24, col sep=comma]{\datatable};
        \addlegendentry{BS-sweep};

        \addplot[black, solid, thick, 
                mark=star, mark size={3.0}, mark size={3.0}, mark_style
                ] 
            table [y=cdf, x=ddpg_df_2x128_gamma060_000, col sep=comma]{\datatable};
        \addlegendentry{Vanilla DDPG};

        \addplot[red, solid, thick, 
                mark=diamond, mark size={3.0}, mark size={3.0}, mark_style
                ] 
            table [y=cdf, x=ddpg_2x128_gamma060_000, col sep=comma]{\datatablenancyvarsigma};
        \addlegendentry{Proposed};
        
        \addplot[green, solid, thick, 
                mark=diamond, mark size={3.0}, mark size={3.0}, mark_style
                ] 
            table [y=cdf, x=ddpg_2x128_gamma060_000, col sep=comma]{\datatablenancybsknownratecdf};
        \addlegendentry{BSOracle};
        
        \addplot[blue, solid, thick, 
                mark=diamond, mark size={3.0}, mark size={3.0}, mark_style
                ] 
            table [y=cdf, x=ddpg_2x128_gamma060_000, col sep=comma]{\datatablenancyangleknownratecdf};
        \addlegendentry{AngleOracle};
    \end{axis}
\end{tikzpicture}
        }
        \caption{$N_t = 16 \times 16$}
        \label{fig:ratecdf_03ue_bs16x16}
    \end{subfigure}%
    \caption{CDF of observed rates for $N_{UE} = 3$.}
    \label{fig:ratecdf_03ue}
\end{figure*}

\begin{figure*}[t]
    \centering
    \begin{subfigure}{.33\linewidth}
        \resizebox{\linewidth}{!}{
            \pgfplotstableread[col sep = comma]{./data/00_Nancy/data_05UE_BS04x04_rateCdf_Baselines_NANZ.csv}\datatable
            \pgfplotstableread[col sep = comma]{./data/02_psuedo_act/data_05UE_BS04x04_rateCdfv1.csv}\tabpsuedoactvone
            \pgfplotstableread[col sep = comma]{./data/00_Nancy/data_05UE_BS04x04_rateCdfNANZ-BS-noisyaction21.csv}\datatablenancyvarsigma
            \pgfplotstableread[col sep = comma]{./data/00_Nancy/data_05UE_BS04x04_rateCdfNANZ-BSKnown.csv}\datatablenancybsknownratecdf
            \pgfplotstableread[col sep = comma]{./data/00_Nancy/data_05UE_BS04x04_rateCdfNANZ-AngleKnown.csv}\datatablenancyangleknownratecdf
            \tikzstyle{mark_style} = [mark size={3.0}, mark repeat=20, mark phase=1]
\begin{tikzpicture}[thick,scale=0.8]
    \begin{axis}[
        width=8cm,
        height=6cm,
        xmin=-1.0,
        xmax=25,
        ymin=-0.1,
        ymax=+1.1,
        grid=major,
        xlabel={Rate (bits/sec/Hz)},
        ylabel={CDF},
        xlabel style={at={(0.50,0.05)}},
        ylabel style={at={(0.06,0.50)}},
        ytick={0.0,0.2,...,1.0},
        legend pos=south east,
        legend cell align={left},
        legend style={fill opacity=0.6, draw opacity=1.0, text opacity=1.0, font=\small}
        ]
        
        \addplot[black, solid, thick, mark=triangle, mark size={3.0}, mark_style
                ] 
            table [y=cdf, x=rand, col sep=comma]{\datatable};
        \addlegendentry{Random};

        \addplot[black, solid, thick, mark=pentagon, mark size={3.0}, mark_style
                ] 
            table [y=cdf, x=greedy, col sep=comma]{\datatable};
        \addlegendentry{Oracle};

        \addplot[black, solid, thick, 
                mark=diamond, mark size={3.0}, mark size={3.0}, mark_style
                ] 
            table [y=cdf, x=sweep24, col sep=comma]{\datatable};
        \addlegendentry{BS-sweep};

        \addplot[black, solid, thick, 
                mark=star, mark size={3.0}, mark size={3.0}, mark_style
                ] 
            table [y=cdf, x=ddpg_df_2x128_gamma060_000, col sep=comma]{\datatable};
        \addlegendentry{Vanilla DDPG};

        \addplot[red, solid, thick, 
                mark=diamond, mark size={3.0}, mark size={3.0}, mark_style
                ] 
            table [y=cdf, x=ddpg_2x128_gamma060_000, col sep=comma]{\datatablenancyvarsigma};
        \addlegendentry{Proposed};
        
        \addplot[green, solid, thick, 
                mark=diamond, mark size={3.0}, mark size={3.0}, mark_style
                ] 
            table [y=cdf, x=ddpg_2x128_gamma060_000, col sep=comma]{\datatablenancybsknownratecdf};
        \addlegendentry{BSOracle};
        
        \addplot[blue, solid, thick, 
                mark=diamond, mark size={3.0}, mark size={3.0}, mark_style
                ] 
            table [y=cdf, x=ddpg_2x128_gamma060_000, col sep=comma]{\datatablenancyangleknownratecdf};
        \addlegendentry{AngleOracle};
    \end{axis}
\end{tikzpicture}
        }
        \caption{$N_t = 4 \times 4$}
        \label{fig:ratecdf_05ue_bs04x04}
    \end{subfigure}%
    \begin{subfigure}{.33\linewidth}
        \resizebox{\linewidth}{!}{
            \pgfplotstableread[col sep = comma]{./data/00_Nancy/data_05UE_BS08x08_rateCdf_Baselines_NANZ.csv}\datatable
            \pgfplotstableread[col sep = comma]{./data/02_psuedo_act/data_05UE_BS08x08_rateCdfv1.csv}\tabpsuedoactvone
            \pgfplotstableread[col sep = comma]{./data/00_Nancy/data_05UE_BS08x08_rateCdfNANZ-BS-noisyaction21.csv}\datatablenancyvarsigma
            \pgfplotstableread[col sep = comma]{./data/00_Nancy/data_05UE_BS08x08_rateCdfNANZ-BSKnown.csv}\datatablenancybsknownratecdf
            \pgfplotstableread[col sep = comma]{./data/00_Nancy/data_05UE_BS08x08_rateCdfNANZ-AngleKnown.csv}\datatablenancyangleknownratecdf
            \pgfplotsset{ignore legend}
            \tikzstyle{mark_style} = [mark size={3.0}, mark repeat=20, mark phase=1]
\begin{tikzpicture}[thick,scale=0.8]
    \begin{axis}[
        width=8cm,
        height=6cm,
        xmin=-1.0,
        xmax=25,
        ymin=-0.1,
        ymax=+1.1,
        grid=major,
        xlabel={Rate (bits/sec/Hz)},
        ylabel={CDF},
        xlabel style={at={(0.50,0.05)}},
        ylabel style={at={(0.06,0.50)}},
        ytick={0.0,0.2,...,1.0},
        legend pos=south east,
        legend cell align={left},
        legend style={fill opacity=0.6, draw opacity=1.0, text opacity=1.0, font=\small}
        ]
        
        \addplot[black, solid, thick, mark=triangle, mark size={3.0}, mark_style
                ] 
            table [y=cdf, x=rand, col sep=comma]{\datatable};
        \addlegendentry{Random};

        \addplot[black, solid, thick, mark=pentagon, mark size={3.0}, mark_style
                ] 
            table [y=cdf, x=greedy, col sep=comma]{\datatable};
        \addlegendentry{Oracle};

        \addplot[black, solid, thick, 
                mark=diamond, mark size={3.0}, mark size={3.0}, mark_style
                ] 
            table [y=cdf, x=sweep24, col sep=comma]{\datatable};
        \addlegendentry{BS-sweep};

        \addplot[black, solid, thick, 
                mark=star, mark size={3.0}, mark size={3.0}, mark_style
                ] 
            table [y=cdf, x=ddpg_df_2x128_gamma060_000, col sep=comma]{\datatable};
        \addlegendentry{Vanilla DDPG};

        \addplot[red, solid, thick, 
                mark=diamond, mark size={3.0}, mark size={3.0}, mark_style
                ] 
            table [y=cdf, x=ddpg_2x128_gamma060_000, col sep=comma]{\datatablenancyvarsigma};
        \addlegendentry{Proposed};
        
        \addplot[green, solid, thick, 
                mark=diamond, mark size={3.0}, mark size={3.0}, mark_style
                ] 
            table [y=cdf, x=ddpg_2x128_gamma060_000, col sep=comma]{\datatablenancybsknownratecdf};
        \addlegendentry{BSOracle};
        
        \addplot[blue, solid, thick, 
                mark=diamond, mark size={3.0}, mark size={3.0}, mark_style
                ] 
            table [y=cdf, x=ddpg_2x128_gamma060_000, col sep=comma]{\datatablenancyangleknownratecdf};
        \addlegendentry{AngleOracle};
    \end{axis}
\end{tikzpicture}
        }
        
        \caption{$N_t = 8 \times 8$}
        \label{fig:ratecdf_05ue_bs08x08}
    \end{subfigure}%
    \begin{subfigure}{.33\linewidth}
        \resizebox{\linewidth}{!}{
            \pgfplotstableread[col sep = comma]{./data/00_Nancy/data_05UE_BS16x16_rateCdf_Baselines_NANZ.csv}\datatable
            \pgfplotstableread[col sep = comma]{./data/02_psuedo_act/data_05UE_BS16x16_rateCdfv1.csv}\tabpsuedoactvone
            
            \pgfplotstableread[col sep = comma]{./data/00_Nancy/data_05UE_BS16x16_rateCdfNANZ-BS-noisyaction21.csv}\datatablenancyvarsigma
            \pgfplotstableread[col sep = comma]{./data/00_Nancy/data_05UE_BS16x16_rateCdfNANZ-BSKnown.csv}\datatablenancybsknownratecdf
            \pgfplotstableread[col sep = comma]{./data/00_Nancy/data_05UE_BS16x16_rateCdfNANZ-BSKnown.csv}\datatablenancybsknownratecdf
            \pgfplotstableread[col sep = comma]{./data/00_Nancy/data_05UE_BS16x16_rateCdfNANZ-AngleKnown.csv}\datatablenancyangleknownratecdf
            \pgfplotsset{ignore legend}
            \tikzstyle{mark_style} = [mark size={3.0}, mark repeat=20, mark phase=1]
\begin{tikzpicture}[thick,scale=0.8]
    \begin{axis}[
        width=8cm,
        height=6cm,
        xmin=-1.0,
        xmax=25,
        ymin=-0.1,
        ymax=+1.1,
        grid=major,
        xlabel={Rate (bits/sec/Hz)},
        ylabel={CDF},
        xlabel style={at={(0.50,0.05)}},
        ylabel style={at={(0.06,0.50)}},
        ytick={0.0,0.2,...,1.0},
        legend pos=south east,
        legend cell align={left},
        legend style={fill opacity=0.6, draw opacity=1.0, text opacity=1.0, font=\small}
        ]
        
        \addplot[black, solid, thick, mark=triangle, mark size={3.0}, mark_style
                ] 
            table [y=cdf, x=rand, col sep=comma]{\datatable};
        \addlegendentry{Random};

        \addplot[black, solid, thick, mark=pentagon, mark size={3.0}, mark_style
                ] 
            table [y=cdf, x=greedy, col sep=comma]{\datatable};
        \addlegendentry{Oracle};

        \addplot[black, solid, thick, 
                mark=diamond, mark size={3.0}, mark size={3.0}, mark_style
                ] 
            table [y=cdf, x=sweep24, col sep=comma]{\datatable};
        \addlegendentry{BS-sweep};

        \addplot[black, solid, thick, 
                mark=star, mark size={3.0}, mark size={3.0}, mark_style
                ] 
            table [y=cdf, x=ddpg_df_2x128_gamma060_000, col sep=comma]{\datatable};
        \addlegendentry{Vanilla DDPG};

        \addplot[red, solid, thick, 
                mark=diamond, mark size={3.0}, mark size={3.0}, mark_style
                ] 
            table [y=cdf, x=ddpg_2x128_gamma060_000, col sep=comma]{\datatablenancyvarsigma};
        \addlegendentry{Proposed};
        
        \addplot[green, solid, thick, 
                mark=diamond, mark size={3.0}, mark size={3.0}, mark_style
                ] 
            table [y=cdf, x=ddpg_2x128_gamma060_000, col sep=comma]{\datatablenancybsknownratecdf};
        \addlegendentry{BSOracle};
        
        \addplot[blue, solid, thick, 
                mark=diamond, mark size={3.0}, mark size={3.0}, mark_style
                ] 
            table [y=cdf, x=ddpg_2x128_gamma060_000, col sep=comma]{\datatablenancyangleknownratecdf};
        \addlegendentry{AngleOracle};
    \end{axis}
\end{tikzpicture}
        }
        \caption{$N_t = 16 \times 16$}
        \label{fig:ratecdf_05ue_bs16x16}
    \end{subfigure}%
    \caption{CDF of observed rates for $N_{UE} = 5$.}
    \label{fig:ratecdf_05ue}
\end{figure*}
\fi

The cumulative distribution of rate achieved by each of the methods is given in Fig. \ref{fig:ratecdf_03ue} and Fig. \ref{fig:ratecdf_05ue}. For each of the methods, we simulated $10000$ observations ($10$ episodes), and the data aggregated based on these observations are plotted. Note that the trained DRL-agents are used to get the rate distribution. Even though the DRL methods can give high rates compared to the baseline methods, as the dimensions of the problem increases, the gap between the proposed approach and the oracle also increases. For example, with less antenna dimension, the proposed method attains a rate close to AngleOracle/BSOracle methods. In a higher-dimensional problem with more antennas and UEs, even though the rate is better than all other DRL/non-DRL based methods, the gap between the proposed method and the Oracle/partially-Oracle methods gives rise to a scope of improvement and is open for further research. 

We can see that, in all cases, the proposed DRL-based blind beam alignment method can provide a better data rate than the beam sweeping method. In the cases where the number of transmit antennas at base stations is small, we can observe that the proposed method can perform close to the Oracle. However, as the number of transmit antennas increases, the gap between the proposed method and Oracle increases. It should be noted that as the number of transmit antennas increases, the beam that can be formed gets narrower and for a given transmit power, better SINR will be delivered at the receiver. Even though the proposed neural network can predict the approximate location of UE from the observed RSSI values, these predictions are susceptible to the noises in the measured RSSI value. When $N_t$ is small, the error in location estimation will not affect the SINR because of the broader beamwidth. With a broader beam, the $\mu BS$ can cover more angular area and hence the requirement for precise angles for beam alignment can be relaxed. But as the number of antenna elements in a $\mu BS$ increases, the beam width gets narrower and hence covering less area in the angular domain. Precise knowledge of location can help in improving the beam alignment accuracy when the number of antenna elements increases. This is the reason for the widening of the gap between the oracle and the proposed method as the number of antenna elements increases. The gap between the proposed method and the AngleOracle method also increases with the increase in the number of antennas for the same reason. However, the proposed method outperforms the other popular traditional as well as machine learning based methods with a very high margin.

It should also be noted that the proposed approach does not have any assumptions on either the UE-BS placements or on the channel properties. This enables the proposed method to be equally applicable to both ground-based as well as aerial UEs/BSs networks.
	\section{Concluding Remarks}
In this work, a deep reinforcement learning-based method is proposed for blind beamforming in mmWave communication systems with a multi-BS multi-UE scenario. The numerical experiments conducted show that the proposed method not only increases the average sum rate of UE four times when compared to the traditional beam-sweeping technique in the case of small antenna arrays but also reaches the same rate as the Oracle. Furthermore, the average sum rate of the proposed method is significantly better than the corresponding rate values of traditional methods like BS-sweep or the DRL based methods like Vanilla DDPG even in the case of larger antenna arrays. This improved performance is achieved with almost no overhead to the system as existing signals in the cellular network can also be used as beacon signal feature extraction. The proposed neural network architecture for handling action spaces which are a mix of discrete and continuous actions is the key part of the improvement in performance. Even though we showed results only with DDPG, the proposed neural architecture is agnostic to the policy gradient method and can be used with any other actor-critic methods where the action is a mix of continuous as well as discrete domains. Further, the proposed neural architecture is also not limited in its utility to only the beamforming problem but can be used in any problem where the action space is a mix of continuous and discrete actions.

 	\bibliographystyle{IEEEtran}
    \bibliography{library.bib}
\end{document}